\documentclass[3p]{elsarticle_nc}
\biboptions{sort&compress}

\usepackage{amsmath,amssymb}
\usepackage{graphicx}
\usepackage{bm}
\usepackage[thinlines]{easybmat}

\newcommand{\fp}[2]{\frac{\partial #1}{\partial #2}} 
\renewcommand{\vec}[1]{\underline{#1}} 
\newcommand{\mat}[1]{\bm{#1}} 
\newcommand{\matelem}[1]{\left[#1\right]} 
\renewcommand{\eqref}[1]{Eq.\,(\ref{#1})} 
\newcommand{\T}{\intercal} 

\renewcommand{\deg}{\mathrm{deg}}
\newcommand{\Ang}{\textup{\AA}}

\begin{document}

\begin{frontmatter}

    \title{The synergy between protein positioning and DNA elasticity: energy
    minimization of protein-decorated DNA minicircles}
    \journal{Nicolas Clauvelin}

    \begin{abstract}
    The binding of proteins onto DNA contributes to the shaping and packaging
    of genome as well as to the expression of specific genetic messages.
    With a view to understanding the interplay between the presence of
    proteins and the deformation of DNA involved in such processes, we
    developed a new method to minimize the elastic energy of DNA fragments at
    the mesoscale level.
    Our method makes it possible to obtain the optimal pathways of
    protein-decorated DNA molecules for which the terminal base pairs are
    spatially constrained.
    We focus in this work on the deformations induced by selected
    architectural proteins on circular DNA.
    We report the energy landscapes of DNA minicircles subjected to different
    levels of torsional stress and containing one or two proteins as functions
    of the chain length and spacing between the proteins.
    Our results reveal cooperation between the elasticity of the double helix
    and the structural distortions of DNA induced by bound proteins.
    We find that the imposed mechanical stress influences the placement of
    proteins on DNA and that, the proteins, in turn, modulate the mechanical
    stress and thereby broadcast their presence along DNA.
    \end{abstract}

    \author[biomaps]{Nicolas Clauvelin}
    \ead{clauvelin@biomaps.rutgers.edu}

    \author[ccb]{Wilma K. Olson}

    \address[biomaps]{BioMaPS Institute for Quantitative Biology, Rutgers, the
    State University of New Jersey, Piscataway, NJ, USA}
    \address[ccb]{BioMaPS Institute for Quantitative Biology and
    Department of Chemistry and Chemical Biology, Rutgers, the State
    University of New Jersey, Piscataway, NJ, USA}


\end{frontmatter}


\section{Introduction}
\label{sec:intro}
The organization of long genomes in the confined spaces of a cell requires
special facilitating mechanisms.
A variety of architectural proteins play key roles in these processes.
Some of these proteins help to compact the DNA by introducing sharp turns
along its pathway while others bring distant parts of the chain molecule into
close proximity.
The histone-like HU protein from
\emph{Escherichia coli} strain U93 and the structurally related Hbb protein
from \emph{Borrelia burgdorferi} induce some of the largest known deformations
of DNA double-helical structure---including global bends in excess of
$160~\mathrm{deg}$, appreciable untwisting, and accompanying dislocations of
the helical axis~\cite{Swinger:Flexible-DNA-bending-in-H:2003,
Swinger:IHF-and-HU:-flexible-arch:2004,
Swinger:Structure-based-Analysis-:2007, Sagi:Modulation-of-DNA-Conform:2004,
Mouw:Shaping-the-Borrelia-burg:2007}.
The degree of DNA deformation reflects both the phasing of the protein-induced
structural distortions and the number of nucleotides wrapped on the protein
surface.
Changes in the spacing between deformation sites on DNA, one such deformation
associated with each half of these two dimeric proteins, contribute to chiral
bends, while variation in the electrostatic surface of the proteins perturbs
the degree of association with DNA and the extent of duplex bending.

By contrast, recent studies have highlighted the important role played by
proteins which bridge and scaffold distant DNA sites and induce the formation
of loops.
For example, the \emph{Escherichia coli} H-NS (histone-like nucleoid
structuring) protein forms a superhelical
scaffold~\cite{Arold:H-NS-forms-a-superhelical:2010} to compact DNA and the
\emph{Escherichia coli} terminus-containing factor MatP can bridge two DNA
sites to form a loop~\cite{Dupaigne:Molecular-Basis-for-a-Pro:2012}.
The MatP protein is responsible for the compaction of a specific macrodomain
(Ter) in \emph{Escherichia coli} and is believed to be able to bind sites
located on either separate chromosomes or within one
chromosome~\cite{Dame:Chromosomal-Macrodomains-:2011}.
The detailed influence of MatP on chromosome segregation and cellular division
is yet to be unraveled~\cite{Dame:Chromosomal-Macrodomains-:2011}, but it is
expected that the presence of MatP proteins could significantly alter the
organization and the dynamics of the whole Ter
macrodomain~\cite{Dupaigne:Molecular-Basis-for-a-Pro:2012}.

The composite kinking and wrapping of DNA on the surface of HU resembles,
albeit on a much smaller scale, the packaging of DNA around the histone
octamer of the nucleosome core
particle~\cite{Luger:Crystal-structure-of-the-:1997}, while MatP might serve
as a bacterial analog of the insulator-binding proteins that divide chromatin
into independent functional
domains~\cite{West:Insulators:-many-function:2002}.
Deciphering how these two very different bacterial proteins influence the
structure and properties of DNA provides a first step in understanding how
architectural proteins contribute to the spatial organization of genomes.

With this goal in mind, we have recently developed a new method for the
optimization of DNA structure at the base-pair level.
Our method takes account of the sequence-dependent elasticity of DNA and can
be applied to chain fragments in which the first and last base pairs are
spatially constrained.
Moreover, our approach allows for constraints in
intervening parts of the DNA and thus makes it possible to model the presence
of bound proteins on DNA.
We can accordingly compute the energy landscape for a wide variety of
protein-DNA systems and herein illustrate the effects of HU and Hbb on the
configurations of circular DNA and provide examples of protein-mediated loops
of the type induced by MatP.

The advantage of our method resides in the direct control of the positions and
orientations of the base pairs at the ends of a DNA chain and in the
capability of accounting for the presence of bound proteins.
In addition, the different constraints due to the boundary conditions and the
presence of proteins are directly integrated in the minimization process,
which makes it possible to rely on unconstrained numerical optimization
methods.
Others have derived methods to optimize the energy of elastic deformation for
DNA fragments.
For example, Coleman et
\emph{al}~\cite{Coleman:Theory-of-sequence-depend:2003} developed a method
similar to ours, but their approach require explicit specification of the
forces and moments acting on the terminal base pairs (that is, there is no
direct control on the positions and orientations of the terminal base pairs).
Zhang and Crothers also presented an optimization
method~\cite{Zhang:Statistical-Mechanics-of-:2003}, albeit limited to circular
DNA. Their method only considers angular deformations within the double helix
and takes the boundary conditions into account through Lagrange multipliers.

We start with a brief description of our method
(Section~\ref{sec:energy_minim}) and then present our results for the effect
of different proteins on circular DNA.
The various appendices contain the details of the calculations.

\section{Energy minimization of spatially constrained, protein-decorated DNA}
\label{sec:energy_minim}
We present in this section a novel method for minimizing the elastic energy of
a collection of DNA base-pairs.
Our method accounts for the sequence-dependent elasticity of DNA and can be
applied to a DNA fragment in which the first and last base pairs are spatially
constrained.
Moreover, our approach makes it possible to constrain parts of the DNA in
order to model the presence of bound proteins.
We describe our method in a symbolic fashion for readability and present full
details of the calculations in the appendices.

\subsection{Geometry of a collection of base pairs}
We consider a collection of $N$ rigid base pairs and for the $i$-th base pair
we denote $\vec{x}^i$ its origin and $\mat{d}^i$ the matrix containing the
axes of the base-pair frame organized as column vectors. 
The segmented curve defined by the base-pair origins $\vec{x}^i$ is referred
to as the collection centerline and can be interpreted as a discrete
double-helical axis.

Here and in the rest of this paper vector symbols are underlined and matrices
are represented with bold symbols.
Superscripted Roman letters ($i, j, \dots$) are used to index base pairs and
base-pair steps.

\subsubsection{Base-pair step parameters}
The geometry of a collection of base pairs is traditionally described in terms
of the base-pair step parameters (or step parameters for
short)~\cite{Dickerson:Definitions-and-nomenclat:1989}.
These parameters describe the relative arrangement of successive base pairs
and are denoted for the $i$-th step $\vec{p}^i=\left(\theta^i_1, \theta^i_2,
\theta^i_3, \rho^i_1, \rho^i_2,
\rho^i_3\right)$~\cite{Coleman:Theory-of-sequence-depend:2003} and referred
to, respectively, as tilt, roll, twist, shift, slide and rise. The first three
parameters correspond to three angles describing the relative orientation of
base-pair frames $\mat{d}^i$ and $\mat{d}^{i+1}$ and the last three parameters
are the components of the step joining vector
$\vec{r}^i=\vec{x}^{i+1}-\vec{x}^i$ expressed in a particular frame referred
to as the step frame (see~\ref{app:bp_step_geometry} for more details
and~\cite{Hassan:The-Assessment-of-the-Geo:1995,
Coleman:Theory-of-sequence-depend:2003} for explicit calculation methods).
The set of step parameters for the base-pair collection is a $6(N-1)$ vector
and is denoted $\vec{P}$.

We shall see later that the energy of elastic deformation for a base-pair step
can be written as a quadratic form with respect to the step parameters.
The step parameters, however, are not a convenient representation to express
constraints on the positions and orientations of base-pairs within the
collection.
This is due to the fact that the components of the step joining vector
$\vec{r}^i$ depend on the relative orientation of the two base pairs forming
the step.
In order to circumvent this issue we introduce a different representation of
the geometry of a base-pair collection.

\subsubsection{Base-pair step degrees of freedom}
We now define a new set of variables for each step of the base-pair
collection. These variables are denoted
$\vec{\phi}^i=\left(\psi^i_1,\psi^i_2,\psi^i_3,r^i_1,r^i_2,r^i_3\right)$ for
the $i$-th step and referred to as the step degrees of freedom or step dofs
for short.
Although the variables $\psi^i$ are identical to the angular step parameters
$\theta^i$, we use a different symbol for clarity.
The variables $r^i$ are the components of the step joining vector
$\vec{r}^i=\vec{x}^{i+1}-\vec{x}^i$ expressed with respect to the global
reference frame (a convenient choice for the global reference frame is the
first base-pair frame).
The set of step dofs for the base-pair collection is a $6(N-1)$ vector and is
denoted $\vec{\Phi}$.

The main advantage of this choice of variables is to separate the
representation of the centerline of the base-pair collection from the
orientation of the base-pair frames. This is somewhat equivalent to the
centerline/spin representation of Langer and
Singer~\cite{Langer:Lagrangian-aspects-of-the:1996} for continuous elastic
rods.
We shall see that this representation is particularly convenient to deal with
the end conditions applied to a collection of base pairs.

\subsection{Energy of elastic deformation}
The energy of elastic deformation for a DNA base-pair collection is defined as
the sum of the energy of elastic deformation for each step.
For the $i$-th step, this energy is given by the following quadratic form:
\begin{equation}
    E^i
    =
    \frac{1}{2}
    \left(\vec{p}^i-\vec{\overline{p}}^i\right)^\T
    \mat{F}^i
    \left(\vec{p}^i-\vec{\overline{p}}^i\right),
    \label{eqn:single_step_energy}
\end{equation}
where $\vec{\overline{p}}^i$ contains the intrinsic step parameters and
describes the reference configuration of the step (\emph{i.e.}, the
configuration of zero energy, also called the rest state), and $\mat{F}^i$ is
a $6\times6$ matrix containing the elastic moduli associated with the
different modes of deformation.
Both the intrinsic step parameters and the force constant matrix can differ
for each step depending on the DNA sequence.
The elastic energy for the base-pair collection is then given by:
\begin{equation}
    \mathcal{E} = \sum_{i=1}^{N-1}E^i.
    \label{eqn:collection_energy}
\end{equation}

The purpose of our method is to minimize the elastic energy of a base-pair
collection given by~\eqref{eqn:collection_energy} under the constraints
detailed below. In other words, we need to calculate the derivatives of the
elastic energy.
The difficulty stems from the fact that this gradient has to be calculated
with respect to a set of independent variables. This set of independent
variables depends on the end conditions applied to the collection of base
pairs and also on the presence of bound proteins.
As mentioned above, the step parameters are not convenient for dealing with
the end conditions. We therefore calculate the gradient with respect to the
step dofs.
We first consider the trivial case of an unconstrained collection
(\emph{i.e.}, a collection with no imposed end conditions and no bound
proteins) and we then show how to account for specific end conditions and the
presence of bound proteins.

Note that, our method does not require the elastic energy of a step to be a
quadratic form with respect to the step parameters. It is also possible to
include coupling between successive steps in the expression of the total
elastic energy of the collection of base pairs.

\subsection{Elastic energy gradient for a free base-pair collection}
The case of a base-pair collection free of any end conditions and without
bound proteins is trivial in the sense that the optimization leads to the
reference configuration.
We will use this gradient for a free collection as the starting point of our
method.

The gradient of the elastic energy for a single step is obtained directly
from~\eqref{eqn:single_step_energy}:
\begin{equation}
    \fp{E^i}{\vec{p}^i} =
    \mat{F}^i_s \left(\vec{p}^i-\vec{\overline{p}}^i\right),
\end{equation}
where we introduce the symmetrized force constant matrix $\mat{F}^i_s$ defined
as $\mat{F}^i_s=(\mat{F}^i+{\mat{F}^i}^\T)/2$.
It follows that the variation of the elastic energy for the complete
collection is given by:
\begin{equation}
    \delta\mathcal{E}
    =
    {\fp{\mathcal{E}}{\vec{P}}}^\T\delta\vec{P}
    =
    \sum_{i=1}^{N-1}{\fp{E^i}{\vec{p}^i}}^\T\delta\vec{p}^i
    =
    \sum_{i=1}^{N-1}
    \left(\mat{F}^i_s \left(\vec{p}^i-\vec{\overline{p}}^i\right)\right)^\T
    \delta\vec{p}^i.
\end{equation}

In order to obtain the gradient with respect to the step dofs we introduce the
Jacobian matrix $\mat{J}_{\vec{\Phi}}$ defined as:
\begin{equation}
    \delta\vec{P}
    =
    \fp{\vec{P}}{\vec{\Phi}}
    \delta\vec{\Phi}
    =
    \mat{J}_{\vec{\Phi}}
    \delta\vec{\Phi}.
    \label{eqn:bp_collection_jacobian}
\end{equation}
It follows that:
\begin{equation}
    \delta\mathcal{E}
    =
    \left(\mat{J}_{\vec{\Phi}}^\T\fp{\mathcal{E}}{\vec{P}}\right)^\T
    \delta\vec{\Phi},
    \label{eqn:free_collection_energy_variation_dofs}
\end{equation}
which leads to the following expression for the free-collection gradient with
respect to the step dofs:
\begin{equation}
    \fp{\mathcal{E}}{\vec{\Phi}}
    =
    {\mat{J}_{\vec{\Phi}}}^\T\fp{\mathcal{E}}{\vec{P}}.
    \label{eqn:free_collection_gradient}
\end{equation}
The details of the calculation for the matrix $\mat{J}_{\vec{\Phi}}$ are given
in~\ref{app:bp_collection_jacobian_matrix}.

\subsection{End conditions}
We now consider the case of a base-pair collection with end conditions, that
is, both the end-to-end vector and the orientation between the first and last
base pairs are fixed.
Our results can easily be generalized to other types of end conditions.

The end-to-end vector for the collection of base pairs is given by:
\begin{equation}
    \vec{x}^N-\vec{x}^1 = \sum_{i=1}^{N-1}\vec{r}^i.
\end{equation}
If the end-to-end vector is imposed, it follows that:
\begin{equation}
    \sum_{i=1}^{N-1}\delta\vec{r}^i = \vec{0},
\end{equation}
and we can therefore write:
\begin{equation}
    \delta\vec{r}^{N-1} = -\sum_{i=1}^{N-2}\delta\vec{r}^i.
    \label{eqn:end_conditions_eed}
\end{equation}

The end-to-end rotation corresponds to the orientation between the first and
last base pairs and is given by:
\begin{equation}
    {\mat{d}^1}^\T\mat{d}^N
    =
   \mat{\mathcal{D}}^{(1,N)}
    =
    \prod_{i=1}^{N-1}\mat{D}^i(\vec{\psi}^i),
\end{equation}
where $\mat{\mathcal{D}}^{(i,j)}$ denotes the rotation matrix between the
$i$-th and $j$-th base pairs, and $\mat{D}^i$ is the rotation matrix for the
$i$-th step which is completely parametrized by the angular step dofs
$\vec{\psi}^i=\left(\psi^i_1,\psi^i_2,\psi^i_3\right)$
(refer to~\ref{app:step_rotation}, \ref{app:collection_rotations} and the
reference~\cite{Coleman:Theory-of-sequence-depend:2003} for more details).
If the end-to-end rotation is imposed we have the constraint:
\begin{equation}
    \delta\mat{\mathcal{D}}^{(1,N)}
    =
    \delta\left(\prod_{i=1}^{N-1}\mat{D}^i(\vec{\psi}^i)\right)
    =
    \mat{0}.
    \label{eqn:eer_constraint}
\end{equation}
This constraint can be written as a condition on $\delta\vec{\psi}^{N-1}$,
that is:
\begin{equation}
    \delta\vec{\psi}^{N-1}
    =
    -\sum_{i=1}^{N-2}\mat{K}^i\delta\vec{\psi}^i,
    \label{eqn:end_conditions_eer}
\end{equation}
where the details of the matrix $\mat{K}^i$ are given
in~\ref{app:end_conditions} (see~\eqref{eqn:matrix_K_definition}).

It follows from the conditions given by~\eqref{eqn:end_conditions_eed}
and~\eqref{eqn:end_conditions_eer} that the variation of the step dofs for the
last step can be expressed as:
\begin{equation}
    \delta\vec{\phi}^{N-1}
    =
    -\sum_{i=1}^{N-1}\mat{B}^i\delta\vec{\phi}^i,
\end{equation}
where, once again, the details of the matrix $\mat{B}^i$ are given
in~\ref{app:end_conditions} (see~\eqref{eqn:end_conditions_B_matrix}).
This result means that the step dofs $\vec{\phi}^{N-1}$ are not independent
variables. In other words, the end conditions reduce the number of independent
step dofs. Note that we choose to express the last step dofs as the
non-independent variables but any other step could have been chosen.
We now denote $\vec{\hat{\Phi}}$ the set of independent step dofs and we have
the relation $\vec{\hat{\Phi}}\subset\vec{\Phi}$.
We then define the Jacobian matrix $\mat{\hat{J}}$ such that:
\begin{equation}
    \delta{\vec{\Phi}}
    =
    \mat{\hat{J}}
    \delta{\vec{\hat{\Phi}}}.
\end{equation}
The dimensions of this matrix depend on the precise details of the end
conditions (although the number of rows is always $6(N-1)$) and the details of
its expression are given in~\ref{app:end_conditions}.

The variation of the elastic energy for a collection of base pairs subjected
to imposed end conditions can be written as:
\begin{equation}
    \delta\mathcal{E}
    =
    \left(
        \left(\mat{J}_{\vec{\Phi}}\mat{\hat{J}}\right)^\T
        \fp{\mathcal{E}}{\vec{P}}
    \right)^\T
    \delta\vec{\hat{\Phi}}.
    \label{eqn:end_conditions_energy_variation_dofs}
\end{equation}
It follows that the gradient is:
\begin{equation}
    \fp{\mathcal{E}}{\vec{\hat{\Phi}}}
    =
    \left(\mat{J}_{\vec{\Phi}} \mat{\hat{J}}\right)^\T
    \fp{\mathcal{E}}{\vec{P}}.
    \label{eqn:end_conditions_gradient}
\end{equation}
Note that this gradient corresponds to the gradient with respect to the
independent step dofs and accounts for the end conditions.
This is one advantage of our method: the end conditions are directly accounted
for. Hence, we transform a constrained optimization problem into an
unconstrained one, which makes the numerical implementation simpler and more
robust.

\subsection{Bound-protein constraint}
In order to study protein-decorated DNA we need to account for the presence of
bound proteins on the double helix.
We model the binding of proteins on the DNA by considering the step parameters
of the binding domain as \emph{frozen}, that is, these step parameters are
imposed and cannot change. For example, the frozen step parameters can be
extracted from high-resolution crystal structures of protein-DNA complexes
with the help of the 3DNA software~\cite{Lu:3DNA:-a-versatile-integra:2008}.

We now consider that the $k$-th step in the base-pair collection is frozen,
that is, the step parameters $\vec{p}^k$ are imposed (the following results
can be easily generalized to an arbitrary number of frozen steps).
It follows directly that the angular step dofs are constant and we therefore
have $\delta\vec{\psi}^k=\vec{0}$. The translational step dofs, however, are
not constant: any changes in the angular steps dofs $\vec{\psi}^j$ with $j<k$
will change the orientation of the joining vector $\vec{r}^k$ of the frozen
step.
Indeed, we show in~\ref{app:frozen_steps} that the variations of the
translational step dofs of the frozen step can be written as:
\begin{equation}
    \delta\vec{r}^k
    =
    \sum_{j=1}^{k-1}
    \mat{W}^{j,k}\delta\vec{\psi}^j.
\end{equation}
We therefore obtain for the variation of the frozen step dofs:
\begin{equation}
    \delta\vec{\phi}^k
    =
    \sum_{j=1}^{k-1}
    \mat{C}^{j,k}\delta\vec{\phi}^j,
\end{equation}
where the details of the matrices $\mat{W}^{j,k}$ and $\mat{C}^{j,k}$ are
given in~\ref{app:frozen_steps} (see~\eqref{eqn:frozen_jacobian_matrix_C}).

Similar to the set of independent step dofs $\vec{\hat{\Phi}}$ introduced for
the treatment of the end conditions, we introduce a new set of independent
step dofs $\vec{\tilde{\Phi}}$ which corresponds to the set of non-frozen step
dofs among the set $\vec{\hat{\Phi}}$, that is, we have
$\vec{\tilde{\Phi}}\subset\vec{\hat{\Phi}}\subset\vec{\Phi}$. We define the
matrix $\mat{\tilde{J}}$ as the Jacobian (see~~\ref{app:frozen_steps} for
details):
\begin{equation}
    \delta\vec{\hat{\Phi}}
    =
    \mat{\tilde{J}}
    \delta\vec{\tilde{\Phi}}.
    \label{eqn:frozen_jacobian}
\end{equation}
We obtain for the variation of the energy:
\begin{equation}
    \delta\mathcal{E}
    =
    \left(
        \left(
            \mat{\tilde{J}}_{\vec{\Phi}}
            \mat{\hat{J}}
            \mat{\tilde{J}}
        \right)^\T
        \fp{\mathcal{E}}{\vec{\tilde{P}}}
    \right)^\T
    \delta\vec{\tilde{\Phi}},
    \label{eqn:frozen_energy_variation_dofs}
\end{equation}
and we have for the gradient:
\begin{equation}
    \fp{\mathcal{E}}{\vec{\tilde{\Phi}}}
    =
    \left(\mat{\tilde{J}}_{\vec{\Phi}}\mat{\hat{J}}\mat{\tilde{J}}\right)^\T
    \fp{\mathcal{E}}{\vec{\tilde{P}}}.
    \label{eqn:frozen_gradient}
\end{equation}
In the above expression, $\vec{\tilde{P}}$ denotes the set of
non-frozen step parameters and $\mat{\tilde{J}}_{\vec{\Phi}}$ is the Jacobian
matrix defined as:
\begin{equation}
    \mat{\tilde{J}}_{\vec{\Phi}}=\fp{\vec{\tilde{P}}}{\vec{\Phi}}.
\end{equation}
This matrix is directly obtained from $\mat{J}_{\vec{\Phi}}$ by removing the
rows associated with the frozen step parameters.

The gradient obtained with~\eqref{eqn:frozen_gradient} corresponds to the
gradient of the elastic energy of the base-pair collection
(\eqref{eqn:collection_energy}) with respect to the set of non-frozen
independent step dofs.

\subsection{Force field}
We introduced in the expression of the elastic energy of a base-pair step
(\eqref{eqn:single_step_energy}) two quantities which depend on the sequence
of the DNA base-pair collection: the intrinsic step parameters
$\vec{\overline{p}}^i$ and the force constants matrix $\mat{F}^i$. These
quantities constitute the force field of our theory. The intrinsic step
parameters describe the rest states of the base-pair steps, while the force
constants are the stiffnesses of the steps.

In this work, we use a uniform ideal force field (\emph{i.e.}, not depending
on the details of the DNA sequence). The intrinsic step parameters for this
force field are given by $\vec{\overline{p}}=\left(0~\deg, 0~\deg,
34.2857~\deg, 0~\Ang, 0~\Ang, 3.4~\Ang\right)$, which correspond to a B-DNA
like straight rest state and a helical repeat of $10.5$ bp. We do not consider
coupling between the different modes of deformation and, hence, the force
constants matrix is diagonal and reads:
\begin{equation}
    \mathrm{diag}\left(\mat{F}^{\mathrm{ideal}}\right)
    =
    \left(
        0.0427,
        0.0427,
        0.0597,
        20,
        20,
        20
    \right).
\end{equation}
The units for the the first three diagonal entries are $k_{B}T.\deg^{-2}$ and
$k_{B}T.\Ang^{-2}$ for the last three entries. This force field can be
considered as quasi-inextensible due to the high force constants associated
with the translational step parameters. In the limit of a infinitely long
base-pair collection we can calculate the persistence lengths corresponding to
the force field and we find that the bending and twisting persistence lengths
are $47.7~\mathrm{nm}$ and $66.6~\mathrm{nm}$, respectively.

\subsection{Protein binding procedure}
As explained earlier, we model the presence of proteins on DNA by setting the
base-pair step parameters of the binding domain to imposed values. These
imposed values can be extracted from high-resolution crystal structures of
protein-DNA complexes or can be chosen to represent an ideal protein-DNA
binding system.

In this study, we use an iterative procedure to bind a protein onto DNA
progressively. For example, we consider a protein with a binding domain of
$n_p$ steps to be bound starting at the $k$-th step, that is, the steps $k$ to
$k+n_p-1$ are bound (and, hence, frozen). The iterative procedure consists in
adjusting the step parameters of the binding domain using the following
\emph{linear ramp} function:
\begin{equation}
    \vec{p}^j(\lambda)
    =
    (1-\lambda)\vec{p}^j_{\mathrm{free}}
    +
    \lambda\vec{p}^j_{\mathrm{protein}},
    \forall j \in \left[k,k+n_p-1\right],
    \label{eqn:binding_ramp}
\end{equation}
where $\vec{p}^j_{\mathrm{free}}$ denote the step parameters of the
protein-free DNA (in our case they will correspond to the step parameters of a
portion of a naked-DNA minicircle) and $\vec{p}^j_{\mathrm{protein}}$ are the
step parameters of the DNA found in the protein-DNA complex. The parameter
$\lambda\in[0,1]$ is the ramp parameter. Our procedure starts with $\lambda=0$
and we minimize the DNA base-pair collection while gradually increasing the
value of the ramp parameter until $\lambda=1$. As an outcome we obtain an
optimized DNA base-pair collection in which a part of DNA is \emph{shaped} as
if the protein were bound to it.
This numerical procedure is not meant to convey the physical process of
proteins binding to DNA. It is designed to set regions of DNA to the states
found in protein-DNA complexes.
We also would like to point out that the binding ramp can be tuned in order to
improve the robustness of the method. For example, the linear terms can be
replaced by quadratic expressions. Such modifications might be needed to avoid
instabilities or to control changes in the total twist while ramping the step
parameters in the case of binding domains with large deformations.

\subsection{Minicircle topology}
The topology of a DNA minicircle is characterized by its linking number
$\mathrm{Lk}$~\cite{Gauss:Carl-Friedrich-Gauss-Werk:1867}, writhing number
$\mathrm{Wr}$~\cite{White:Calculation-of-the-twist-:1986,
Pohl:The-Self-Linking-Number-o:1968}, and total twist
$\mathrm{Tw}$~\cite{White:Calculation-of-the-twist-:1986}.
The linking number of a minicircle, an integer, describes the entanglement of
the minicircle centerline and the curve traced out by one of the edges of the
base-pair frames (see~\cite{Clauvelin:Characterization-of-the-G:2012} for
detailed explanations and computational methods).
In particular, for a planar minicircle the linking number corresponds to the
number of turns the double helix makes.
For a minicircle of $N$ bp, the \emph{relaxed} linking number $\mathrm{Lk}^0$
is given by the integer nearest to $N/10.5$, where $10.5$ is the
assumed helical repeat of DNA.
We introduce the difference between the actual and relaxed linking numbers of
a minicircle as $\Delta\mathrm{Lk}=\mathrm{Lk}-\mathrm{Lk}^0$. Because a DNA
minicircle is covalently closed, its linking number, and hence
$\Delta\mathrm{Lk}$, are constant and are not altered by the deformation of
the double helix.

The linking number of a minicircle is always equal to the sum of its writhing
number and total twist, $\mathrm{Lk}=\mathrm{Tw}+\mathrm{Wr}$.
The writhing number characterizes the global folding and non-chiral
distortions of the minicircle centerline, while the total twist measures the
twisting or
\emph{twist density} of the base pairs around the centerline.
Like the linking number, the writhing number and total twist can be directly
obtained from the minicircle base-pair frames as explained
in~\cite{Clauvelin:Characterization-of-the-G:2012}.
The invariance of the linking number implies that when the minicircle is
deformed, the total twist and the writhing number are redistributed. The total
twist is directly related to the torsional stress within the double helix,
while the writhing number depends on the curvature of the centerline, albeit
in a non-trivial way.

\subsection{Protein-free DNA minicircles}
In order to study protein-decorated DNA minicircles we first need to generate
protein-free minicircles. For a minicircle of $N$ bp, we first build a set of
step parameters using the following formula for the $i$-th step:
\begin{equation}
    \vec{p}^i
    =
    \left(
        \Delta \sin\left(\theta_{3}^\star i\right),
        \Delta \cos\left(\theta_{3}^\star i\right),
        \theta_{3}^\star,
        0,
        0,
        \overline{\rho}_{3}
    \right),
\end{equation}
where $\Delta=2\pi/N$ is the bending angle between the planes of successive
base pairs and $\overline{\rho}_{3}$ denotes the intrinsic value of the rise
step parameter within our force field. The parameter $\theta_{3}^\star$
corresponds to the local twist density in the minicircle and can be adjusted
to generate over- or undertwisted configurations. In order to enforce the
covalent closure of the minicircle, we add at the end of the base-pair
collection an additional base-pair identical to the first one. We then
minimize the energy of the base-pair collection under the imposed end-to-end
vector and rotation (which are both null in the present case). The outcome of
the minimization is an optimized DNA minicircle with an imposed
$\Delta\mathrm{Lk}$.

Here we initially focus on planar, protein-free minicircles of lengths ranging
from 63 bp to 105 bp (\emph{i.e.}, from 6 to 10 helical repeats). Planar
minicircles always have a writhing number equal to zero, which means that the
total twist is equal to the linking number.
Because we require the naked minicircles to be planar there is an upper bound
on the value of $\left|\mathrm{Tw}\right|$ (and, hence, of
$\left|\Delta\mathrm{Lk}\right|$). This limiting value is related to the
Michell-Zajac instability~\cite{Zajac:Stability-of-two-planar-l:1962,
Goriely:Twisted-Elastic-Rings-and:2006}, and for larger values of
$\left|\mathrm{Tw}\right|$ the planar circular configurations are no longer
stable.
With our ideal force field, for a given chain length we can always generate
optimized minicircles with $\Delta\mathrm{Lk}=0$ and, depending on the chain
length, $\Delta\mathrm{Lk}=-1$ or $\Delta\mathrm{Lk}=+1$ (for a few specific
chain lengths the three values are possible; see~\ref{app:zajac_instability}
for more details).
In other words, we always compute optimized protein-free minicircles with
$\Delta\mathrm{Lk}=0$ and in addition, we also generate optimized minicircles
with $\Delta\mathrm{Lk}=-1$ and/or $\Delta\mathrm{Lk}=+1$.
These different minicircles are then used with the binding ramp
(\eqref{eqn:binding_ramp}) to model the presence of bound proteins.

\section{Results}
\label{sec:results}
We first apply our minimization method to the optimization of DNA minicircles
on which one or two proteins are bound.
We focus on two distinct proteins: the abundant histone-like HU protein and
the structurally related Hbb protein.
These two proteins have a common fold and share some structural similarities,
such as the fact that they both associate as dimers and both introduce
significant localized bends in the double helix.
We used the 3DNA software~\cite{Lu:3DNA:-a-versatile-integra:2008} to extract
the step parameters of the DNA found in known crystal
complexes~\cite{Swinger:Flexible-DNA-bending-in-H:2003,
Mouw:Shaping-the-Borrelia-burg:2007} (Protein Data Bank files 1P71 and 2NP2
for HU and Hbb, respectively). The lengths of the extracted binding domains
are 17 bp for HU and 35 bp for Hbb. Within the scope of our work, we think of
the Hbb-bound DNA as an extreme model of HU-induced distortion.

The net bending angle introduced by the HU dimer in the selected structure is
roughly $135~\mathrm{deg}$, and that by Hbb is $160~\mathrm{deg}$ (these
angles correspond to the angles between the normals of the first and last base
pairs).
Both proteins also slightly underwind the double helix. That is, the total
twist across the DNA bound to each protein is less than the total twist of an
undeformed DNA fragment of same length (for HU $\Delta\mathrm{Tw}\simeq
-0.18~\mathrm{turns}$ and for Hbb $\Delta\mathrm{Tw}\simeq
-0.1~\mathrm{turns}$).

\subsection{Minicircles with a single bound protein}
We first performed a series of optimizations to study how the addition of an
HU or Hbb dimer affects a DNA minicircle.
We focused on chain lengths of 63 bp to 105 bp and considered relaxed
minicircles ($\Delta\mathrm{Lk}=0$) as well as over- or undertwisted
minicircles ($\Delta\mathrm{Lk}=\pm1$).

\begin{figure}[htb]
        \center
        \includegraphics[width=.55\textwidth]{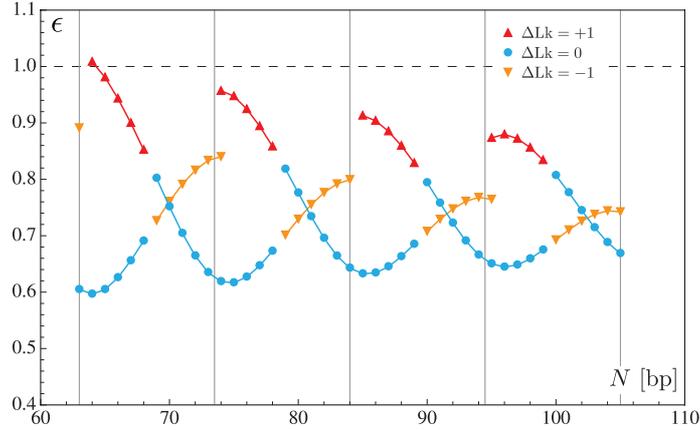}
        \caption{Chain-length dependence of the relative step energy
        $\epsilon$ for minicircles with a single HU dimer. The blue circles
        correspond to the energies of relaxed minicircles and span the whole
        range of chain lengths. The red and orange triangles represent the
        energies of over- and underwound minicircles, respectively. The
        horizontal dashed line ($\epsilon=1$) corresponds to the energy of a
        base-pair step in a naked minicircle of the same length. The vertical
        lines indicate the chain lengths equal to integral numbers of helical
        repeats.}
        \label{fig:HU_chain_length}
    \end{figure}

\begin{figure}[htb]
        \center
        \includegraphics[width=.55\textwidth]{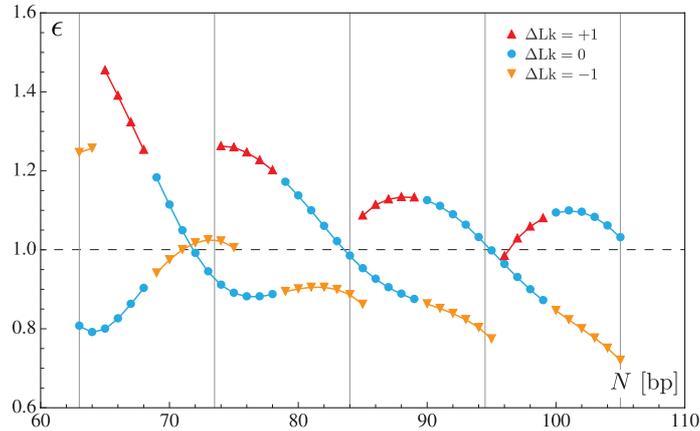}
        \caption{Chain-length dependence of the relative step energy
        $\epsilon$ for minicircles with a single Hbb dimer. The blue circles
        correspond to the energies of relaxed minicircles and span the whole
        range of chain lengths. The red and orange triangles represent the
        energies of over- and underwound minicircles, respectively. The
        horizontal dashed line ($\epsilon=1$) corresponds to the energy of a
        base-pair step in a naked minicircle of the same length. The vertical
        lines indicate the chain lengths equal to integral numbers of helical
        repeats.}
        \label{fig:Hbb_chain_length}
    \end{figure}

In order to characterize these optimized configurations we define a relative
step energy $\epsilon$ as:
\begin{equation}
     \epsilon=\frac{N}{N_\mathrm{free}}\frac{\mathcal{E}}{\mathcal{E}^0},
\end{equation}
where $\mathcal{E}$ is the energy of the optimized minicircle with the bound
dimer, $\mathcal{E}^0$ is the energy of the optimized protein-free minicircle,
$N$ denotes the minicircle chain length, and $N_\mathrm{free}$ the
protein-free chain length.
In other words, $\epsilon$ is the ratio of the energy per base-pair step for
the minicircle with a single protein versus that for the naked minicircle. We
recall that our elastic energy for a minicircle accounts for protein-free
steps only. The results for HU are presented in Fig.~\ref{fig:HU_chain_length}
and those for Hbb in Fig.~\ref{fig:Hbb_chain_length}.

We first remark that the presence of an HU dimer almost always reduces the
energy of the protein-free DNA compared to that of the naked minicircle (as
shown by the values of the relative step energy $\epsilon$ less than one in
Fig.~\ref{fig:HU_chain_length}).
On the other hand, the addition of an Hbb protein has a mixed effect on the
relative step energy depending on the chain length and the value of
$\Delta\mathrm{Lk}$ (Fig.~\ref{fig:Hbb_chain_length}).
A possible explanation lies in the fact that the binding domain of the Hbb
dimer is twice as long as that for the HU dimer. Thus, the length of
protein-free DNA is less for minicircles with an Hbb protein, although the
boundary conditions are comparable to those of HU. In other words, the
deformation required to satisfy the boundary conditions is larger in
minicircles with an Hbb dimer, and, hence, the energy is higher.
This argument also accounts for the differences in the chain-length dependence
of the relative step energy for minicircles containing an HU dimer versus
those with an Hbb dimer.
As shown in Fig.~\ref{fig:HU_chain_length}, for the HU dimer the chain-length
dependence of the relative step energy follows a damped oscillatory pattern
(the period is roughly equal to the assumed 10.5-bp DNA helical repeat).
By contrast, there is no clear periodicity in the Hbb results
(Fig.~\ref{fig:Hbb_chain_length}).
This suggests, that within the present range of chain lengths, a minicircle
with a single Hbb dimer is more \emph{constrained} than one with an HU
protein.

The other interesting feature found in these results is the influence of the
linking number and the total twist on the relative step energy.
In the case of the HU dimer, relaxed minicircles have, in most cases, a lower
energy than under- or overwound minicircles. It is only for chain lengths
close to half-integral numbers of helical repeats that the energy is lower for
underwound minicircles.
On the other hand, underwound minicircles of 79 bp or greater bearing an Hbb
dimer are consistently lower in energy than relaxed or overwound minicircles.
As mentioned earlier, both dimers slightly underwind the double helix, which
means that their presence on a minicircle induces a redistribution of the
torsional stress.
Note that, this redistribution may entail the conversion of part of the
torsional stress into bending deformations as a consequence of the changes in
the total twist and the writhing number.
This also explains why overwound minicircles lead to higher energies: the
large difference in the torsional stress between the naked DNA (prior to the
addition of a protein) and the bound DNA causes higher deformations in the
remaining part of the minicircle.
In addition, the torsional stress in naked DNA depends on the chain length:
for chain lengths close to integral numbers of helical repeats, the torsional
stress is lower in relaxed minicircles than in underwound minicircles, but for
chain lengths near half-integral numbers of helical repeats the situation is
reversed.

Our results show that the torsional stress in DNA may influence the
recruitment of HU and Hbb dimers and might act as a
\emph{control mechanism} for the presence of such proteins.
For example, a relaxed minicircle of 94 bp is more likely to be bound to an HU
than an Hbb dimer (assuming that the binding affinity of both dimers are
comparable), whereas an underwound minicircle of the same length is more
likely to take up an Hbb dimer.
Although our results are obtained on covalently closed DNA minicircles (and,
hence, under a topological constraint), it is reasonable to think that similar
effects will take place on torsionally constrained linear DNA fragments (for
example, the anchoring conditions could be achieved by large protein
assemblies or magnetic/optical tweezers).

\subsection{Minicircles with two bound proteins}
\begin{figure}[htb]
        \center
        \includegraphics[width=\textwidth]{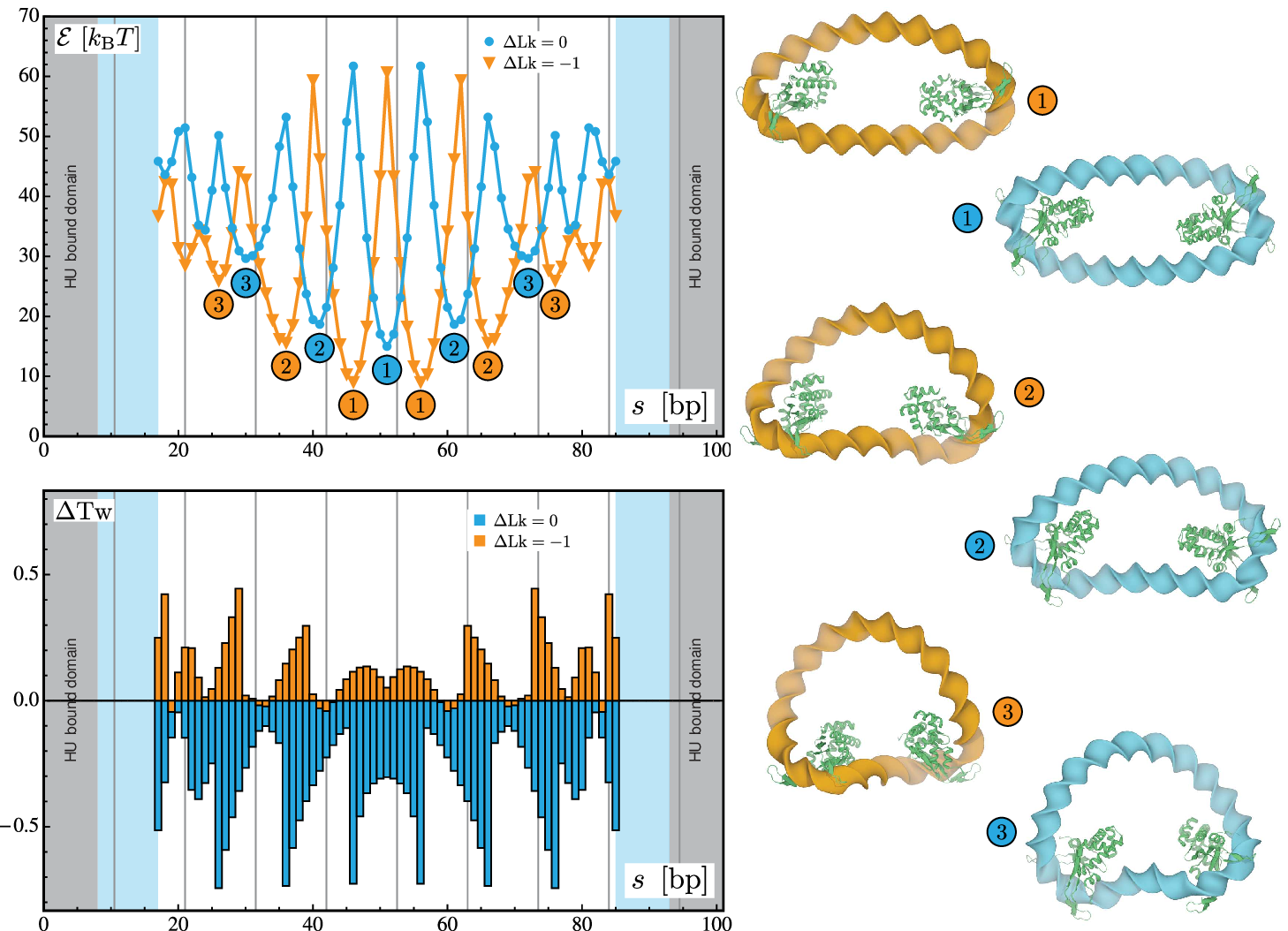}
        \caption{Optimization results for relaxed and underwound minicircles
        of 100 bp with two HU dimers.
        The two plots on the left represent (top) the optimized energy and
        (bottom) the changes in the total twist as functions of the
        center-to-center spacing $s$ between the two proteins.
        In both plots, the gray areas denote the binding domain of the first
        HU protein and the light blue areas the binding domain of the second
        protein.
        The vertical lines indicate the chain lengths equal to integral
        numbers of helical repeats.
        The numbers in the energy plot refer to the structures depicted on the
        right in which the HU proteins are represented in green.
        The underwound ($\Delta\mathrm{Lk}=-1$) and relaxed
        ($\Delta\mathrm{Lk}=0$) minicircles are represented in orange and
        blue, respectively.}
        \label{fig:100bp_HUx2_panel}
    \end{figure}

\begin{figure}[htb]
        \center
        \includegraphics[width=\textwidth]{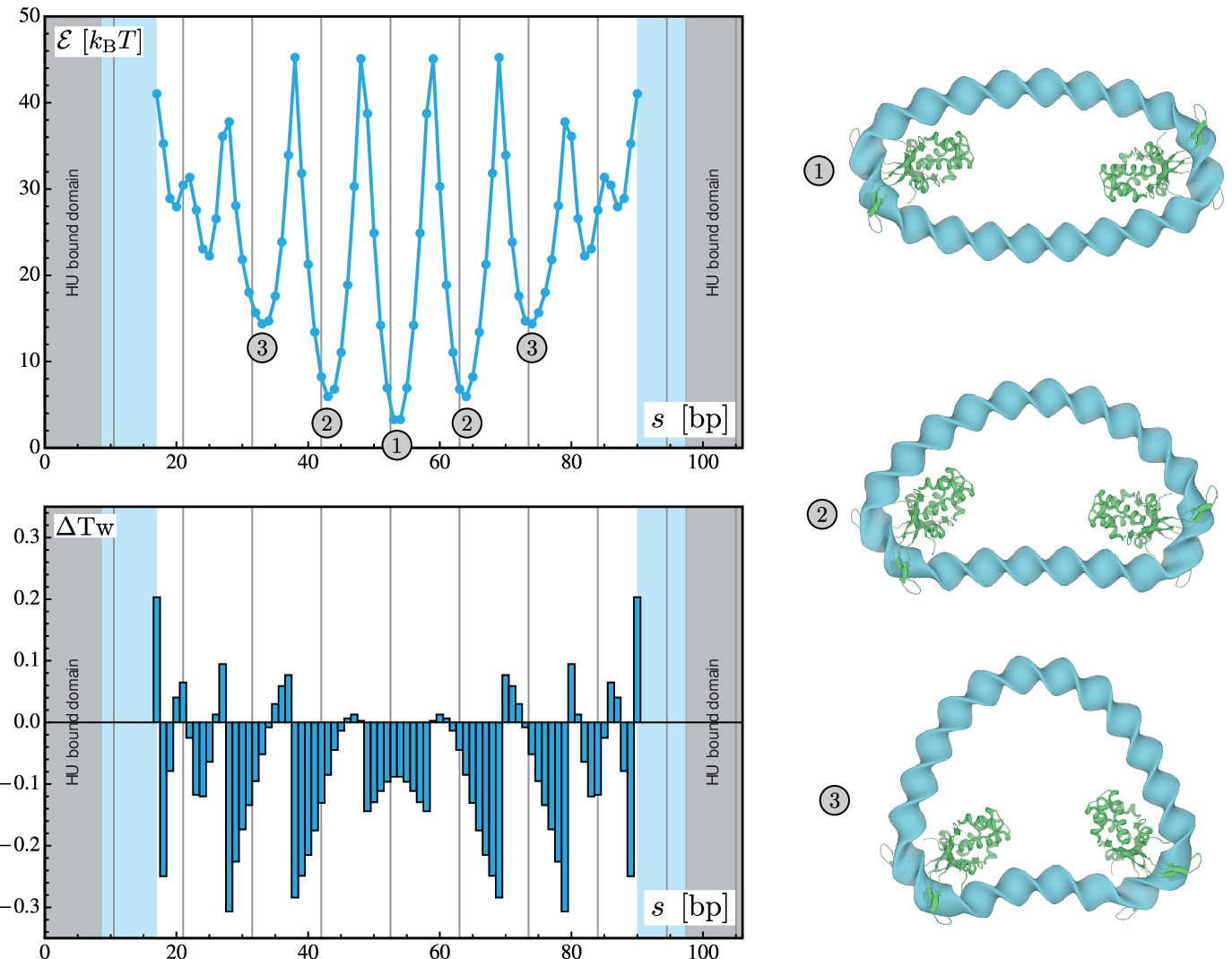}
        \caption{Optimization results for minicircles of 105 bp with two HU
        dimers.
        The two plots on the left represent (top) the optimized energy and
        (bottom) the changes in the total twist as functions of the
        center-to-center spacing $s$ between the two proteins.
        In both plots, the gray areas denote the binding domain of the first
        HU protein and the light blue areas the binding domain of the second
        protein. The vertical lines indicate the chain lengths equal to
        integral numbers of helical repeats.
        The numbers in the energy plot refer to the structures depicted on the
        right in which the HU proteins are represented in green.}
        \label{fig:105bp_HUx2_panel}
    \end{figure}

\begin{figure}[htb]
        \center
        \includegraphics[width=\textwidth]{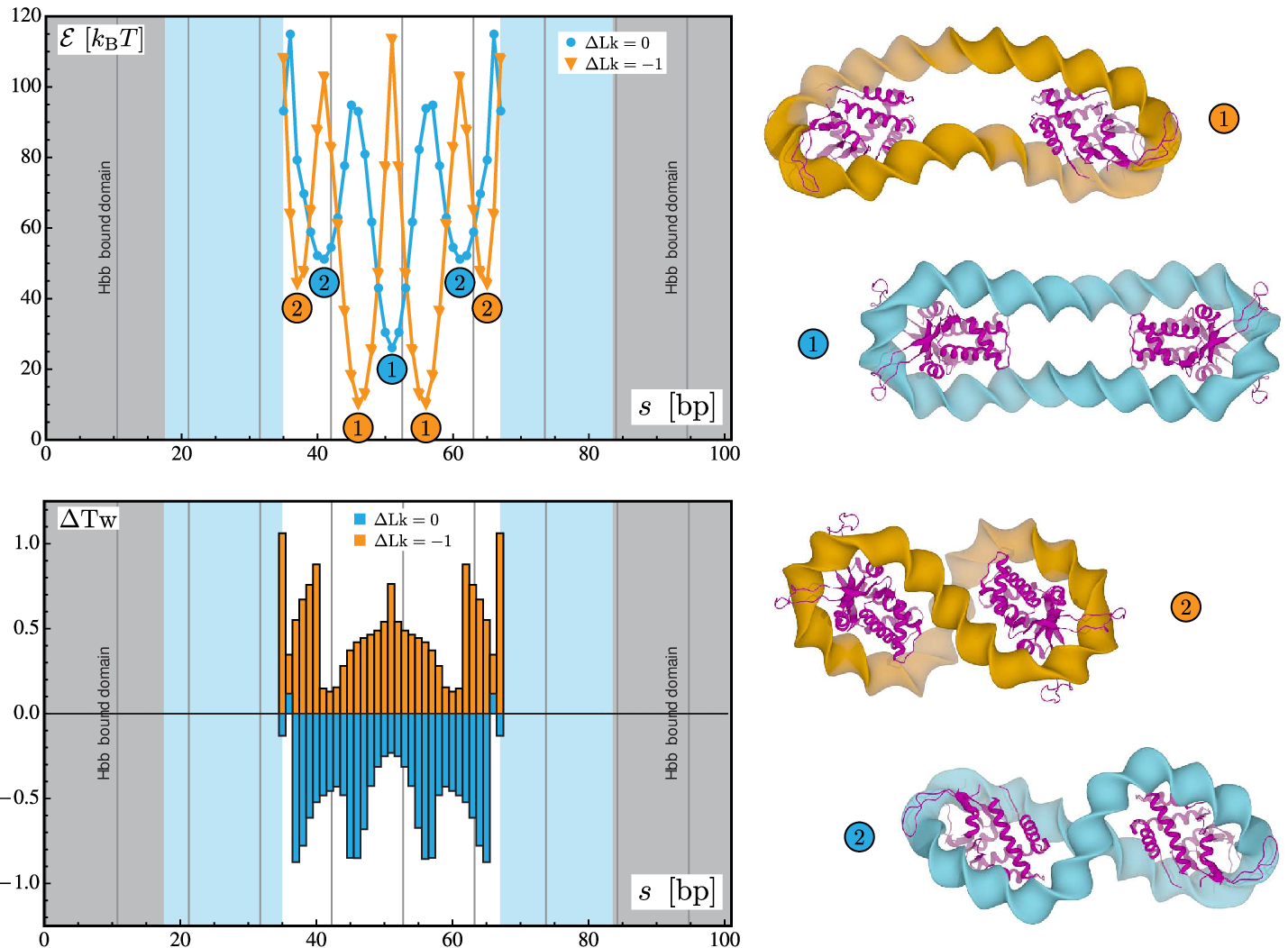}
        \caption{Optimization results for relaxed and underwound minicircles
        of 100 bp with two Hbb dimers.
        The two plots on the left represent (top) the optimized energy and
        (bottom) the changes in the total twist as functions of the
        center-to-center spacing $s$ between the two proteins.
        In both plots, the gray areas denote the binding domain of the first
        Hbb protein and the light blue areas the binding domain of the second
        protein. The vertical lines indicate the chain lengths equal to
        integral numbers of helical repeats.
        The numbers in the energy plot refer to the structures depicted on the
        right in which the Hbb proteins are represented in pink.
        The underwound ($\Delta\mathrm{Lk}=-1$) and relaxed
        ($\Delta\mathrm{Lk}=0$) minicircles are represented in orange and
        blue, respectively.}
        \label{fig:100bp_HBBx2_panel}
    \end{figure}

\begin{figure}[htb]
        \center
        \includegraphics[width=\textwidth]{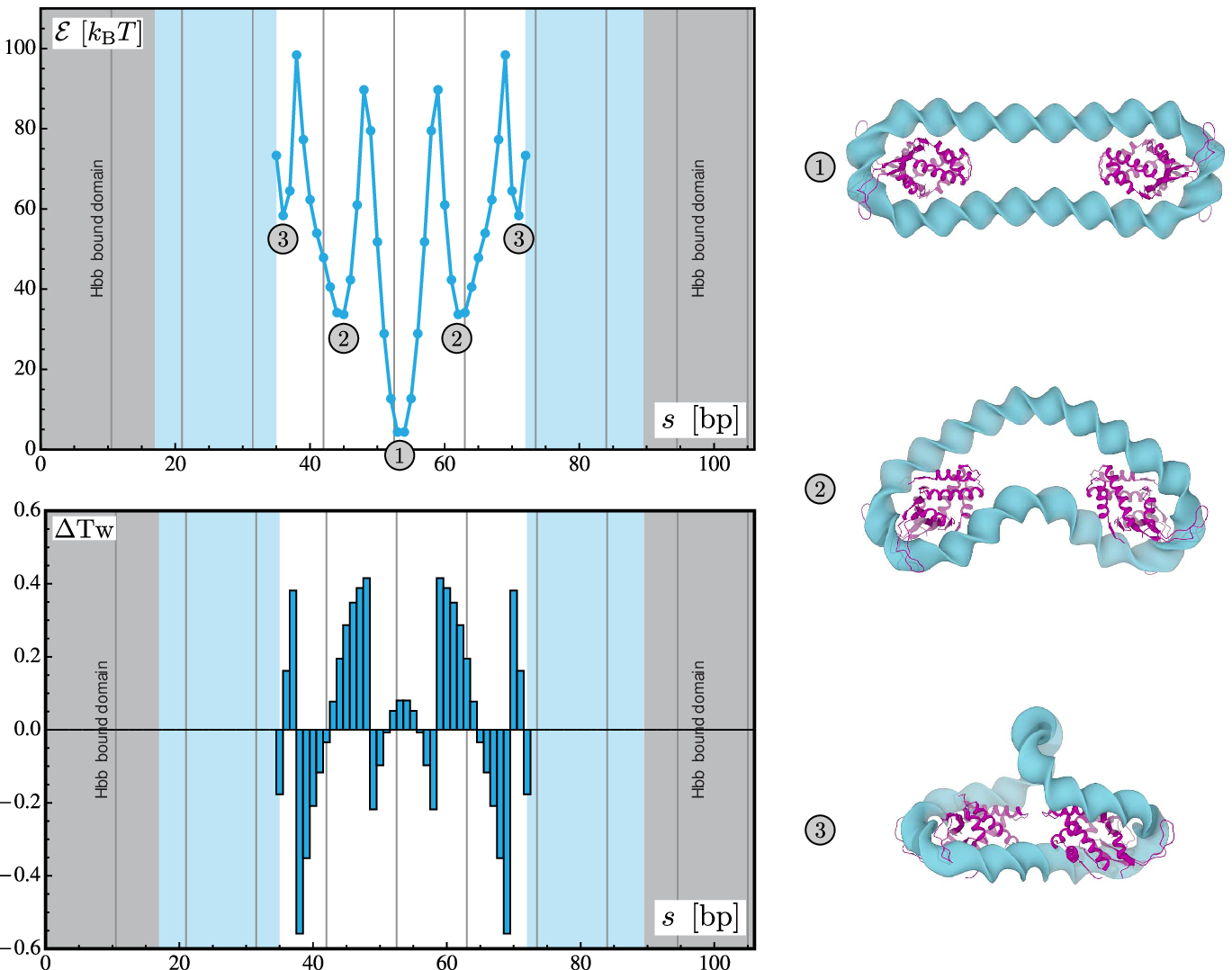}
        \caption{Optimization results for minicircles of 105 bp with two Hbb
        dimers.
        The two plots on the left represent (top) the optimized energy and
        (bottom) the changes in the total twist as functions of the
        center-to-center spacing $s$ between the two proteins.
        In both plots, the gray areas denote the binding domain of the first
        Hbb protein and the light blue areas the binding domain of the second
        protein. The vertical lines indicate the chain lengths equal to
        integral numbers of helical repeats.
        The numbers in the energy plot refer to the structures depicted on the
        right in which the Hbb proteins are represented in pink.}
        \label{fig:105bp_HBBx2_panel}
    \end{figure}

Our second series of optimizations focuses on minicircles of 100 bp and 105 bp
containing two HU or two Hbb dimers.
In addition, for minicircles of 100 bp we consider relaxed
($\Delta\mathrm{Lk}=0$) and underwound ($\Delta\mathrm{Lk}=-1$) molecules.
This choice is motivated by the fact that the difference in energy between a
naked DNA minicircle of 100 bp with $\Delta\mathrm{Lk}=0$ and
$\Delta\mathrm{Lk}=-1$ is only of $1.8~\mathrm{k_BT}$ (the relaxed minicircle
corresponds to the lower energy).
Our computations provide energy landscapes of the protein-bound minicircles as
functions of the spacing between the two binding sites.
These landscapes are directly related to the relative likelihoods of forming
minicircles with two dimers at specific locations. In other words, the
landscapes reveal which dimer positions along the minicircle are more apt to
be occupied in, for example, cyclization experiments.
We report in Figs.~\ref{fig:100bp_HUx2_panel}-\ref{fig:105bp_HBBx2_panel} the
energy landscapes as well as the associated changes in the total twist
$\Delta\mathrm{Tw}$ (with respect to the planar and naked minicircles).

The energy landscapes consist of several local minima separated by high energy
states.
In other words, there are well-defined locations for optimal placements of two
HU or Hbb proteins along the DNA minicircles.
Notice that, only the minima with the lowest energies and, hence, with the
highest Boltzmann weights, are likely to be relevant to the statistical
physics of protein-decorated minicircles.
These minima appear periodically in the landscapes and, as expected, the
period is roughly equal to the assumed DNA helical repeat (10.5 bp).
In addition, the relative locations of the two proteins along the minicircle
appear to affect the torsional stress significantly as evidenced by the
variations in the total twist.
We also notice that the lowest energy configurations are all similar from a
geometric point of view. That is, for both HU and Hbb proteins the globally
optimal configurations are obtained when the two dimers are located at
antipodal or near antipodal sites (as shown by the configurations labeled `1'
in Fig.~\ref{fig:100bp_HUx2_panel} to Fig.~\ref{fig:105bp_HBBx2_panel}).
We note that the minima for the minicircles with two HU dimers are of lower
energy than those for minicircles with two Hbb dimers.
For Hbb-bound minicircles the length of protein-free DNA is shorter than for
HU (for two Hbb dimers, the length of naked DNA is comparable to the size of
the Hbb binding domain).
The Hbb system is therefore highly constrained and leads to larger values in
the energy landscapes and greater changes in the total twist.

The results for the minicircles of 100 bp containing two HU or two Hbb dimers
(see Figs.~\ref{fig:100bp_HUx2_panel} and~\ref{fig:100bp_HBBx2_panel}) show
that, the energy minima are lower for the underwound minicircle
($\Delta\mathrm{Lk}=-1$) than for the relaxed minicircle
($\Delta\mathrm{Lk}=0$).
This is consistent with the results obtained for a minicircle of 100 bp with a
single dimer (see Figs.~\ref{fig:HU_chain_length}
and~\ref{fig:Hbb_chain_length}).
We remark that the variations in the energy landscapes of the relaxed and
underwound minicircles of 100 bp are out of phase. That is, the minima for the
relaxed minicircles correspond to the maxima of the underwound minicircles and
\emph{vice versa}. This \emph{phase shift} between the energy landscapes of
relaxed and underwound minicircles is roughly equal to half the assumed
helical repeat ($\sim 5\!-\!6$ bp).
The changes in the total twist for these minicircles are also similar,
although the magnitude of the change is larger for the minicircles with two
Hbb proteins. For the relaxed minicircles of 100 bp the changes in the total
twist are negative (with respect to the planar, protein-free minicircles),
while for underwound minicircles the changes are positive.
This implies, because of the conservation of the linking number, that the
changes in the writhing number are of opposite sign for relaxed versus
underwound minicircles.
Indeed, as shown in Fig.~\ref{fig:100bp_HBBx2_panel} (configurations labeled
`2'), the minicircles are of different \emph{handedness}.
We also notice that, the pattern in the changes in the total twist for
minicircles of 100 bp with two dimers is similar: the local maxima in the
energy correspond to the highest changes in the total twist, while the local
minima correlate with those of moderate values of $\Delta\mathrm{Tw}$.

For the minicircles of 105 bp containing two HU or Hbb dimers, the energy
landscapes are similar to those obtained for 100-bp minicircles.
In particular, the minima appear periodically (every $\sim 10.5$ bp) and are
separated by high energy configurations.
The minima for the 105-bp chains are lower than those on the minicircles of
100 bp. This difference is due to the fact that in 105-bp minicircles the
torsional stress in the minicircle prior to the addition of the proteins is
lower than that in the 100-bp minicircles.
The main dissimilarity between minicircles of 100 bp and 105 bp resides in the
changes in the total twist.
We have seen that for 100-bp minicircles, the changes in the total twist are
comparable for HU and Hbb dimers.
In the case of the minicircles of 105 bp, however, the values of
$\Delta\mathrm{Tw}$ corresponding to the local minima are of different signs
for the Hbb and HU minicircles.
That is, for the lower energy configurations, the presence of two HU dimers
reduces the torsional stress in the minicircle (see
Fig.~\ref{fig:105bp_HUx2_panel}), while two Hbb dimers increase the torsional
stress (see Fig.~\ref{fig:105bp_HBBx2_panel}).
We also notice that, for these local minima the magnitude of the changes in
the total twist is comparable.

Our study of minicircles with two dimers reveals the interplay between the
elastic properties of DNA and the positioning of proteins on the double-helix.
It appears that this is not about the proteins shaping the double helix nor
the DNA stiffnesses controlling the positioning of proteins. Rather there is
cooperation between the presence of proteins and the elasticity of the double
helix, particularly in the distribution of the torsional stress.
The high degree of contrast in our energy landscapes suggests that, once a
protein is bound to a topologically constrained DNA fragment, the stress in
the double helix will favor specific binding sites for other proteins.
Such an interpretation echoes the recent results about DNA-protein allosteric
effects obtained in single-molecule
experiments~\cite{Kim:Probing-Allostery-Through:2013}.
It is also interesting to notice that the local minima found in our energy
landscapes are always flanked by configurations of comparable energy (up to a
few $\mathrm{k_BT}$).
This suggests that there should be fluctuations in the experimental
measurements of the most likely binding sites of HU and Hbb dimers along DNA
minicircles.
Notice that, our results have been obtained with a minimal model, that is, DNA
is modeled as an isotropic material with standard bending and twisting
stiffnesses and we have focused on a special class of protein geometry which
includes specific deformations on the double helix.
Nevertheless, our approach serves as a proof of concept and paves the way for
more detailed studies about the synergy between DNA deformation and protein
positioning.

\section{Discussion}
\label{sec:discussion}
The minimization procedure introduced in this work facilitates the
investigation of how architectural proteins may contribute to the spatial
organization and genetic processing of DNA.
This new approach gives us direct control of the positions and orientations of
the base pairs at the ends of a DNA chain and allows us to specify the precise
sites of protein uptake and the detailed changes in double-helical structure
brought about by the binding of protein.
Here we illustrate the utility of the method in a study of the elastic energy
landscapes of DNA minicircles decorated by the nonspecific architectural
protein HU and the similarly folded, albeit
site-specific~\cite{Kobryn:Site-specific-DNA-binding:2000}, Hbb protein.
Both proteins associate as dimers and introduce severe bends and untwist their
DNA targets. We consider the Hbb-bound DNA as an extreme example of HU-induced
DNA distortion and thus treat both proteins as nonspecific.
We focus on covalently closed molecules comparable in length to the loops that
are formed by various regulatory proteins and enzymes that bind to
sequentially distant sites on DNA~\cite{Adhya:Multipartite-Genetic-Cont:1989,
Schleif:DNA-Looping:1992} and allow for the uptake of one or two HU or Hbb
dimers on the DNA.
We also consider underwound and overwound minicircles in order to study the
added effects of the torsional stress within the double helix on the
protein-binding landscapes.
We find that the presence of protein has a significant effect on the bending
and twisting deformations in the minicircles, and conversely, that the
torsional stress within DNA prior to the addition of proteins has a strong
effect on the optimal placement of proteins along the minicircles.
For example, we show that a single HU dimer is more likely to bind relaxed
rather than under- or overwound minicircles of most chain lengths between
63-105 bp and that an Hbb dimer binds preferentially to underwound minicircles
of the same lengths.

Our results reveal cooperation between the deformability of the double helix
and the structural distortions of DNA induced by bound proteins.
In the case of minicircles with two HU or two Hbb dimers, the presence of a
first protein strongly influences the locations of the optimal binding sites
of a second protein.
That is, the DNA, through its elastic deformation, acts as a communication
medium between the proteins. In particular, the torsional stress and the
twisting stiffness appear to play a major role in this action at a distance.
The mechanical signaling also provides a rationale for the DNA allostery
reported in recent single-molecules studies of the dissociation of proteins on
DNA chains constrained to full extension by the flow of
solvent~\cite{Kim:Probing-Allostery-Through:2013,
Xu:Modeling-Spatial-Correlat:2013}. In particular, the binding of one protein
on the extended DNA stabilizes or destabilizes the binding of another protein,
even when not in direct contact.
Indeed, this sort of long-range communication, in which the binding of a
ligand in one part of the DNA helix influences (positively or negatively) the
recognition of a different ligand at a remote site, also termed
telestability~\cite{Burd:Transmission-of-stability:1975}, has puzzled DNA
scientists for decades.
Although our results concern a special class of proteins bound to covalently
closed rather than straightened DNA, we observe an interplay between the
elasticity of the double helix and the placement of proteins along DNA that
may hold for extended as well as cyclic chains.
In our case, the most likely placement of a second protein is antipodal to the
first, \emph{i.e.}, as sequentially far apart as possible. Our studies provide
examples of how the mechanical stress in DNA can control the placement of
proteins and how proteins can alter the mechanical stress to broadcast their
presence along DNA.
Our findings also suggest that local modifications of the mechanical
properties of DNA, such as methylation or the occurrence of kinks, could
modulate and possibly repress this type of mechanical signaling.

    \begin{figure}[htb]
        \center
        \includegraphics[width=.85\textwidth]{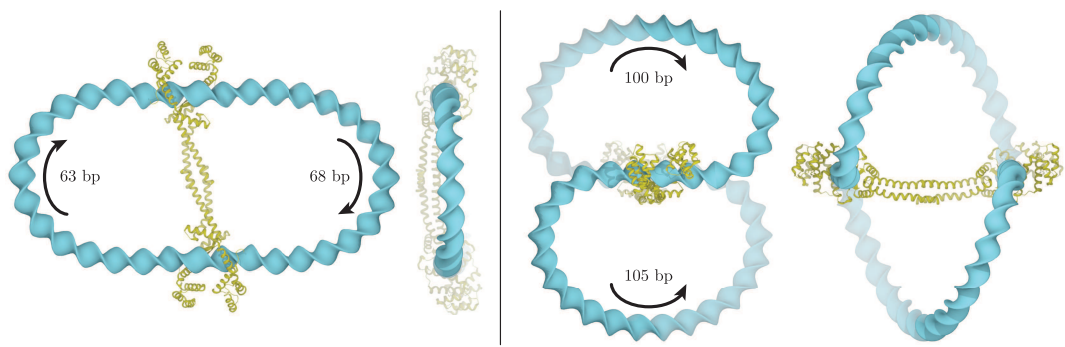}
        \caption{Orthogonal views of optimized loops induced by the
        DNA-bridging protein
        MatP~\cite{Dupaigne:Molecular-Basis-for-a-Pro:2012} on circular DNA.
        The loops depicted on the left side run in anti-parallel directions
        and those on the right side run in parallel directions (note the
        arrows and chain lengths in the images).
        The total chain lengths of the circular DNA on the left and right
        sides are 177 bp and 251 bp, respectively.
        The MatP dimeric protein is represented in yellow.
        The step parameters of the two bound DNA domains have been extracted
        from Protein Data Bank file 3VEA.}
        \label{fig:matp_loops}
    \end{figure}

    \begin{figure}[htb]
        \center
        \includegraphics[width=.95\textwidth]{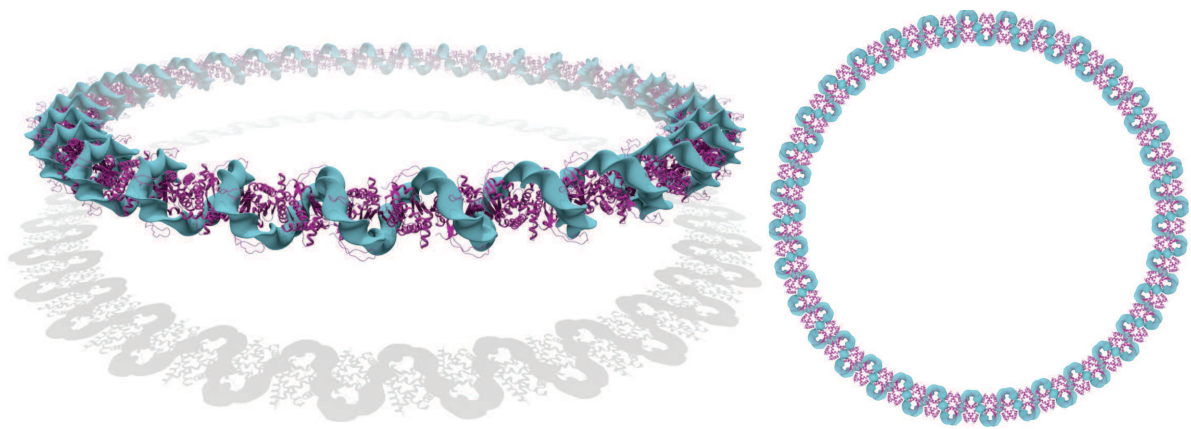}
        \caption{Perspective (left) and orthographic (right) views of an
        optimized DNA plasmid of 1280 bp containing 64 Hbb proteins regularly
        spaced by 5-bp linkers.
        The presence of the Hbb proteins (represented
        in pink) greatly condenses the DNA into a helical \emph{fiber-like}
        structure that folds into a circular configuration.
        The circular
        radius of the resulting superstructure is $31.5~\mathrm{nm}$ and the
        fiber radius is $3~\mathrm{nm}$ (the same DNA plasmid without any
        proteins has a circular radius of $69.2~\mathrm{nm}$ and a
        cross-sectional radius of $1~\mathrm{nm}$).
        The elastic energy stored in each naked-DNA linkers is
        $0.19~\mathrm{k_BT}$, which suggests that the linkers are lowly
        deformed.
        In contrast to the zigzag superhelical structure found in the crystal
        lattice (see Fig.~1c in~\cite{Mouw:Shaping-the-Borrelia-burg:2007}) in
        which the proteins lie on opposite sides of the superhelical axis, the
        proteins form a central core in the optimized structure.
        Notice that, this structure has been obtained with a \emph{reduced}
        Hbb model in which only the fifteen central base pairs of the full
        model described earlier are used.
        As a consequence, the spacing between the proteins is greater in the
        optimized structure than in the pseudo-continous helix found in the
        crystal lattice.}
        \label{fig:Hbb_SuperStructure}
    \end{figure}

Although the minimal elastic energy configurations of long DNA chains are not
necessarily relevant from a statistical physics point of view, our method can
be used to study larger systems, such as long DNA molecules decorated by
multiple proteins.
For example, we show in Fig.~\ref{fig:matp_loops} two examples of multiple
loops induced by the Ter-specific protein
MatP~\cite{Dupaigne:Molecular-Basis-for-a-Pro:2012} on circular DNA, which is
thought to be involved in the condensation of chromosomes in \emph{Escherichia
coli}. We also show in Fig.~\ref{fig:Hbb_SuperStructure} a long DNA plasmid
decorated by 64 closely spaced Hbb proteins.
Our software can also be used together with other tools (for example,
3DNA~\cite{Lu:3DNA:-a-versatile-integra:2008}) to optimize DNA fragments
anchored by proteins.
An interesting application of our approach resides in the development of
software for biomolecular
sculpting~\cite{Touzain:DNA-motifs-that-sculpt-th:2011}, which makes it
possible to study how proteins can bundle and organize DNA.
We are currently working on improving our method to account for additional
types of constraints, such as the treatment of excluded volume.
To date, our software does not check for collisions between DNA and proteins.
Although the proteins collide on some of the minicircles presented in this
work (see Figs.~\ref{fig:100bp_HUx2_panel}-\ref{fig:105bp_HBBx2_panel}), the
collisions only occur in high-energy configurations, in which pairs of
proteins lie immediately next to one another.
We also plan to include more realistic force fields to account, for example,
for the higher deformability of pyrimidine-purine compared to other base-pair
steps and for the flexibility of protein assemblies. This task is facilitated
by the fact that our method already accounts for the sequence-dependent
elasticity of DNA.
Finally, we are now concentrating our efforts on other biomolecular systems,
including the loops mediated by the Lac and Gal repressor proteins and the
relative contributions of protein and DNA flexibility in loop formation.

\bibliographystyle{elsarticle-num}
\bibliography{dna_minim}

\clearpage
\appendix

\section{Base-pair step geometry}
\label{app:bp_step_geometry}
We consider in this appendix and the following ones a collection of $N$ rigid
base pairs and for the $i$-th base pair we denote $\vec{x}^i$ its origin and
$\mat{d}^i$ the matrix containing the axes of the base-pair frame organized as
column vectors. The step parameters of the $i$-th step are denoted $\vec{p}^i$
and the step dofs are denoted $\vec{\phi}^i$. The definitions of the step
parameters are given in~\cite{Dickerson:Definitions-and-nomenclat:1989,
Hassan:The-Assessment-of-the-Geo:1995, Coleman:Theory-of-sequence-depend:2003}
(in particular, the first reference provides expressions for the angular step
parameters in terms of the Euler angles describing the rotation between two
successive base pairs).

Vector symbols are underlined and matrices are represented with bold symbols
and the elements of a vector or a matrix are denoted with square brackets.
Superscripted Greek letters ($\alpha, \beta, \dots$) are used to denote
vector, matrix, or tensor entries and range from 1 to 3. Superscripted Roman
letters ($i, j, \dots$), on the other hand, are used to index base pairs and
base-pair steps.
For example, the $\alpha$-th axis of base-pair frame $\mat{d}^i$ is denoted
$\vec{d}^i_\alpha$, and the $\beta$-th component of that vector is
$\matelem{\vec{d}^i_\alpha}_\beta=\matelem{\mat{d}^i}_{\alpha\beta}$.
We also use the Einstein summation notation on lower indices: for example, the
scalar product of two vectors $\vec{u}$ and $\vec{v}$ is given by
$\matelem{\vec{u}}_\alpha\matelem{\vec{v}}_\alpha$, that is,
$\vec{u}\cdot\vec{v}=\matelem{\vec{u}}_\alpha\matelem{\vec{v}}_\alpha =
\sum_{\alpha=1}^{3}\matelem{\vec{u}}_\alpha\matelem{\vec{v}}_\alpha$.

\subsection{Step rotation}
\label{app:step_rotation}
For a given base-pair step, we denote $\mat{D}^i$ the matrix describing the
rotation between the two successive base-pair frames $\mat{d}^i$ and
$\mat{d}^{i+1}$. This rotation matrix is referred to as the step rotation
matrix and is defined by:
\begin{equation}
    \mat{D}^i(\vec{\psi}^i) = {\mat{d}^i}^\T\mat{d}^{i+1}.
    \label{eqn:step_rotation_matrix}
\end{equation}

One of the main results about the geometry of a base-pair step is the relation
between the changes in the angular step dofs and the changes in the relative
orientation of the base-pair frames. We assume that the base-pair frames are
orthonormal and it follows that an infinitesimal perturbation of a base-pair
frame axis $\vec{d}^i_\alpha$ can be written as
$\delta\vec{d}^i_\alpha=\vec{\chi}^i\times\vec{d}^i_\alpha$.
The changes in the step rotation matrix due to the perturbation of the two
base-pair frames forming the step is therefore given by:
\begin{equation}
    \matelem{\delta\mat{D}^i}_{\alpha\beta}
    =
    \delta\vec{\chi}^i\cdot\vec{d}^{i+1}_\beta\times\vec{d}^{i}_\alpha
    =
    -\matelem{\mat{D}^i}_{\gamma\beta}\epsilon_{\alpha\gamma\zeta}
        \vec{d}^i_{\zeta}\cdot
        \delta\vec{\chi}^i,
\end{equation}
where $\delta\vec{\chi}^i=\vec{\chi}^{i+1}-\vec{\chi}^{i}$ and
$\epsilon_{\alpha\gamma\zeta}$ is the three-dimensional Levi-Civita symbol. We
introduce the vector $\delta\vec{\omega}^i={\mat{d}^i}^\T\delta\vec{\chi}^i$
which is the projection of $\delta\vec{\chi}^i$ in the $i$-th base-pair frame.
It follows that:
\begin{equation}
    \matelem{(\delta\mat{D}^i){\mat{D}^i}^\T}_{\alpha\beta}
    =
    -\epsilon_{\alpha\beta\gamma}\matelem{\delta\vec{\omega}^i}_{\gamma}.
\end{equation}
The term on the left-hand side corresponds to a skew matrix because of the
orthogonality of $\mat{D}^i$ and we introduce the matrix $\mat{\Xi}^i$ such
that:
\begin{equation}
    \matelem{(\delta\mat{D}^i){\mat{D}^i}^\T}_{\alpha\beta}
    =
    \matelem{
        \fp{\mat{D}^i}{\psi^i_\gamma}{\mat{D}^i}^\T
    }_{\alpha\beta}\delta\psi^i_\gamma
    =
    -\epsilon_{\alpha\beta\zeta}\matelem{\mat{\Xi}^i}_{\zeta\gamma}
    \psi^i_\gamma
    \label{eqn:step_Xi_matrix}
\end{equation}
We finally obtain after contracting the Levi-Civita symbols:
\begin{equation}
    \delta\vec{\omega}^i=\mat{\Xi}^i\delta\vec{\psi}^i.
\end{equation}
This expression relates the relative changes in the base-pair frames to the
variations of the angular step dofs. The matrix $\mat{\Xi}^i$ is
invertible~\cite{Coleman:Theory-of-sequence-depend:2003} and we introduce the
matrix $\mat{\Omega}^i={\mat{\Xi}^i}^{-1}$ to obtain:
\begin{equation}
    \delta\vec{\psi}^i
    =
    \mat{\Omega}^i\delta\vec{\omega}^i.
    \label{eqn:step_Omega_matrix}
\end{equation}
The matrix $\mat{\Omega}^i$ can be obtained by direct calculation and its
transpose is identical to the matrix ${\mat{\Gamma}^n}$ given in Eq.~(A.1)
of~\cite{Coleman:Theory-of-sequence-depend:2003}.

\subsection{Step frame}
The step itself can be characterized by introducing a so-called step frame
(referred to as the mid-step frame
in~\cite{Hassan:The-Assessment-of-the-Geo:1995}). This frame is located at
$(\vec{x}^i+\vec{x}^{i+1})/2$ and the matrix containing the step frame axes as
columns, denoted $\mat{s}^i$, is given by:
\begin{equation}
    \mat{s}^i=\mat{d}^i\mat{D}^i_s(\vec{\psi}^i),
    \label{eqn:step_frame_definition}
\end{equation}
where $\mat{D}^i_s$ is a rotation matrix that can be obtained directly from
the angular step dofs (see~\cite{Coleman:Theory-of-sequence-depend:2003} for
more details).

\subsection{Translational step parameters}
The step parameters $\vec{\rho}^i$ are defined as the projection of the step
joining vector $\vec{r}^i=\vec{x}^{i+1}-\vec{x}^i$ on the step frame, that is,
we have:
\begin{equation}
    \vec{\rho}^i={\mat{s}^i}^\T\vec{r}^i.
    \label{eqn:step_parameters_rho_definition}
\end{equation}

The definition given by~\eqref{eqn:step_parameters_rho_definition} shows that
the step parameters $\vec{\rho}^i$ depend on the step dofs $\vec{r}^i$ and on
the angular step dofs $\vec{\psi}^i$. We introduce the matrix $\mat{T}^i$ such
that:
\begin{equation}
    \delta\vec{\rho}^i=\mat{T}^i\delta\vec{r}^i,
\end{equation}
and we have:
\begin{equation}
    \matelem{\mat{T}^i}_{\alpha\beta}=\fp{\rho^i_\alpha}{r^i_\beta}
    =
    \matelem{{\mat{s}^i}^\T}_{\alpha\beta}
    .
    \label{eqn:jacobian_matrix_T}
\end{equation}
We also introduce the matrix $\mat{R}^i$ such that:
\begin{equation}
    \delta\vec{\rho}^i=\mat{R}^i\delta\vec{\psi}^i.
\end{equation}
We have:
\begin{equation}
    \matelem{\mat{R}^i}_{\alpha\beta}
    =
    \fp{\rho^i_\alpha}{\psi^i_\beta}
    =
    \matelem{\fp{{\mat{D}^i_s}^\T}{\psi^i_\beta}\mat{D}^i_s}_{\alpha\gamma}
    \rho^i_\gamma,
\end{equation}
where the matrix appearing in the right-hand side is skew ($\mat{D}^i_s$ is
orthogonal). We denote $\vec{\Lambda}^i_\beta$ its vector representations and
we obtain:
\begin{equation}
    \matelem{\mat{R}^i}_{\alpha\beta}
    =
    \matelem{\vec{\Lambda}^i_\beta\times\vec{\rho}^i}_{\alpha}.
    \label{eqn:jacobian_matrix_R}
\end{equation}
The vectors $\vec{\Lambda}^i_\beta$ can be obtained by direct calculation and
are identical to the ones given in Eqs~(A.2-A.4)
in~\cite{Coleman:Theory-of-sequence-depend:2003}.

\section{Base-pair collection geometry}
\label{app:bp_collection_geometry}
We derive in this appendix the results related to the Jacobian matrix
$\mat{J}_{\vec{\Phi}}$. This Jacobian matrix is defined as
(see~\eqref{eqn:bp_collection_jacobian}):
\begin{equation}
    \mat{J}_{\vec{\Phi}}=\fp{\vec{P}}{\vec{\Phi}},
\end{equation}
where $\vec{P}$ denotes the set of all base-pair step parameters and
$\vec{\Phi}$ the set of all step dofs.

\subsection{Rotations within a base-pair collection}
\label{app:collection_rotations}
We denote $\mat{\mathcal{D}}^{(i,j)}$ the matrix describing the rotation
between the base-pair frames $\mat{d}^i$ and $\mat{d}^j$ ($i<j$). This
rotation matrix is defined as:
\begin{equation}
    \mat{\mathcal{D}}^{(i,j)}
    =
    {\mat{d}^i}^\T\mat{d}^j
    =
    \prod_{k=i}^{j-1}\mat{D}^k(\vec{\psi}^k),
    \label{eqn:base_pairs_rotation_matrix}
\end{equation}
where $\mat{D}^k(\vec{\psi}^k)$ is the step rotation matrix
(see~\eqref{eqn:step_rotation_matrix}).

We want to calculate the derivatives of this product of rotation matrices with
respect to the angular step dofs $\vec{\psi}^k$ ($i\leq k < j$). We
have:
\begin{equation}
    \fp{\mat{\mathcal{D}}^{(i,j)}}{\psi^k_\alpha}
    =
    \mat{D}^i\dots\mat{D}^{k-1}\fp{\mat{D}^k}{\psi^k_\alpha}
        \mat{D}^{k+1}\dots\mat{D}^{j-1},
\end{equation}
which can be rewritten as:
\begin{equation}
    \fp{\mat{\mathcal{D}}^{(i,j)}}{\psi^k_\alpha}
    =
    {\mat{d}^i}^\T\mat{d}^{k}
        \left(\fp{\mat{D}^k}{\psi^k_\alpha}{\mat{D}^k}^\T\right)
    {\mat{d}^k}^\T\mat{d}^{j}.
\end{equation}
The matrix in parentheses on the right-hand side is a skew matrix
(see~\eqref{eqn:step_Xi_matrix}) and we introduce the notation $\left(\partial
\mat{D}^k/\partial\psi^k_\alpha\right){\mat{D}^k}^\T=\mat{P}^k_\alpha$ such
that:
\begin{equation}
    \fp{\mat{\mathcal{D}}^{(i,j)}}{\psi^k_\alpha}
    =
    {\mat{d}^i}^\T\mat{d}^{k}
    \mat{P}^k_\alpha
    {\mat{d}^k}^\T\mat{d}^{j}
    =
    \mat{\mathcal{D}}^{(i,k)}
    \mat{P}^k_\alpha
    \mat{\mathcal{D}}^{(k,j)},
    \label{eqn:rotation_product_derivatives}
\end{equation}
and the transpose is given by:
\begin{equation}
    \left(\fp{\mat{\mathcal{D}}^{(i,j)}}{\psi^k_\alpha}\right)^\T
    =
    {\mat{\mathcal{D}}^{(k,j)}}^\T
    {\mat{P}^k_\alpha}^\T
    {\mat{\mathcal{D}}^{(i,k)}}^\T
    =
    -{\mat{\mathcal{D}}^{(k,j)}}^\T
    {\mat{P}^k_\alpha}
    {\mat{\mathcal{D}}^{(i,k)}}^\T,
\end{equation}
where we used the property ${\mat{P}^k_\alpha}^\T=-\mat{P}^k_\alpha$.

We can use this result to calculate the derivatives of a base-pair frame with
respect to the angular step dofs of the preceding steps. The $i$-th base-pair
frame in the collection is given by:
\begin{equation}
    \mat{d}^i=\mat{d}^1\mat{\mathcal{D}}^{(1,i)}.
\end{equation}
It follows from~\eqref{eqn:rotation_product_derivatives} that:
\begin{equation}
    \fp{\mat{d}^i}{\psi^k_\alpha}
    =
    \left(\mat{d}^{k}\mat{P}^k_\alpha{\mat{d}^k}^\T\right)\mat{d}^i
\end{equation}
The matrix multiplying $\mat{d}^i$ on the right-hand side is a skew matrix and
can therefore be represented by a vector. This vector expresses the
infinitesimal rotation due to changes in the $\vec{\psi}^k$. We define the
vector $\vec{S}^k_\alpha$ as:
\begin{equation}
    \matelem{\vec{S}^k_\alpha}_\beta
    =
    -\frac{1}{2}\epsilon_{\gamma\zeta\beta}
    \matelem{\mat{d}^k\mat{P}^k_\alpha{\mat{d}^k}^\T}_{\gamma\zeta}.
\end{equation}
This expression can be rewritten in a more concise form with the help
of~\eqref{eqn:step_Xi_matrix} ($\mat{P}^k_\alpha$ is the skew matrix obtained
from the $\alpha$-th column of $\mat{\Xi}^k$):
\begin{equation}
    \vec{S}^k_\alpha=\mat{d}^j\vec{\Xi}^k_\alpha,
    \label{eqn:S_vectors_definition}
\end{equation}
where $\vec{\Xi}^k_\alpha$ is the $\alpha$-th column of $\mat{\Xi}^k$. In
other words, $\vec{S}^k_\alpha$ is the expression of the vector
$\vec{\Xi}^k_\alpha$ in the global reference frame and the vector
$\vec{S}^k_\alpha$ describes the infinitesimal rotation associated with a
change in the $\alpha$-th angular step dof of the $k$-th step. We finally
obtain:
\begin{equation}
    \fp{}{\psi^k_\alpha}\matelem{\mat{d}^k}_{\gamma\zeta}
    =
    \fp{}{\psi^k_\alpha}\matelem{\vec{d}^k_\zeta}_{\gamma}
    =
    \matelem{\vec{S}^k_\alpha\times\vec{d}^i_\zeta}_\gamma.
    \label{eqn:frame_derivative_S_vectors}
\end{equation}

\subsection{Jacobian matrix \texorpdfstring{$\mat{J}_{\vec{\Phi}}$}
    {for the base-pair collection}}
\label{app:bp_collection_jacobian_matrix}
We already have two of the three contributions to the Jacobian matrix of the
collection: the matrices $\mat{T}^i$ and $\mat{R}^i$
(see~\eqref{eqn:jacobian_matrix_T} and~\eqref{eqn:jacobian_matrix_R}) express
the dependence of the $\vec{\rho}^i$ on the step dofs $\vec{\psi}^i$ and
$\vec{r}^i$.
The definition of the $\vec{\rho}^i$
(\eqref{eqn:step_parameters_rho_definition}) shows that there is also a
dependence on the angular dofs from the preceding steps in the collection. We
have from~\eqref{eqn:step_parameters_rho_definition}:
\begin{equation}
    \fp{\rho^i_\alpha}{\psi^j_\beta}
    =
    \matelem{
        {\mat{D}^i_s}^\T\left(\fp{\mat{d}^i}{\psi^j_\beta}\right)^\T
    }_{\alpha\gamma}
    r^i_\gamma, \;\mathrm{where}\; j<i
\end{equation}
which, with the help of~\eqref{eqn:frame_derivative_S_vectors}, leads to:
\begin{equation}
    \fp{\rho^i_\alpha}{\psi^j_\beta}
    =
    -\matelem{
        \left({\mat{s}^i}^\T\vec{S}^j_\beta\right)\times\vec{\rho}^i
    }_\alpha
    =
    \matelem{\mat{U}^{j,i}}_{\alpha\beta}.
    \label{eqn:jacobian_matrix_U}
\end{equation}

\begin{figure}[htb]
        \center
        \includegraphics[width=.5\textwidth]{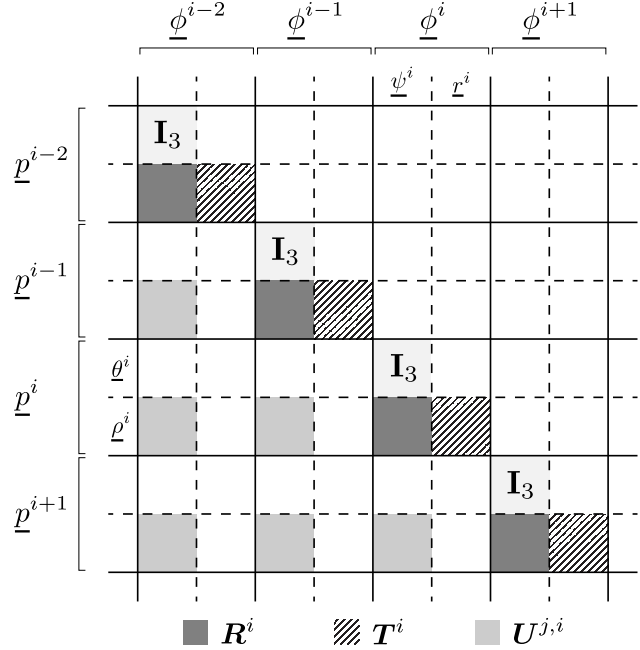}
        \caption{Structure of the matrix $\mat{J}_{\vec{\Phi}}$ for four
        \emph{central} steps $i-2,\dots,i+1$. The vectors $\vec{p}^i$ denote
        the step parameters while the vectors $\vec{\phi}^i$ denote the step
        dofs. The solid lines delimit the $6\times6$ base-pair step block
        matrices, while the dashed lines separate the angular variables from
        the translational variables ($\vec{\theta}^i$ and $\vec{\rho}^i$ for
        the step parameters, and $\vec{\psi}^i$ and $\vec{r}^i$ for the step
        dofs). The white blocks represent null entries.}
        \label{fig:transformation_matrix}
    \end{figure}

We can now write the full variation of the step parameters $\vec{p}^i$ in
terms of the step dofs $\vec{\phi}^k$ $(1\leq k\leq i)$ as:
\begin{equation}
    \delta\vec{p}^i
    =
    \left[\:
        \begin{BMAT}(@,18pt,18pt){c:c}{c:c}
            \mat{\mathrm{I}}_3 & \mat{0} \\
            \mat{R}^i & \mat{T}^i
        \end{BMAT}
    \:\right]
    \delta\vec{\phi}^i
    +
    \sum_{j=1}^{i-1}
    \left[\:
        \begin{BMAT}(@,18pt,18pt){c:c}{c:c}
            \mat{0} & \mat{0}\\
            \mat{U}^{j,i} & \mat{0}
        \end{BMAT}
    \:\right]
    \delta\vec{\phi}^j,
\end{equation}
where $\mat{R}^i$ is defined in~\eqref{eqn:jacobian_matrix_R}, $\mat{T}^i$
in~\eqref{eqn:jacobian_matrix_T}, and $\mat{U}^{j,i}$
in~\eqref{eqn:jacobian_matrix_U}. The matrix $\mat{J}_{\vec{\Phi}}$ is
directly obtained from this expression and its structure is shown in
Fig.~\ref{fig:transformation_matrix}.

\section{End conditions}
\label{app:end_conditions}
We provide in this appendix the details for the calculation of the Jacobian
matrix $\mat{\hat{J}}$. This Jacobian matrix relates the set of
\emph{independent} step dofs to the complete set of step dofs. The number of
independent step dofs depends on the details of the boundary conditions. We
focus here on an imposed end-to-end vector and an imposed end-to-end rotation
but other types of boundary conditions can be treated along the same lines.

The imposed end-to-end vector condition is given
by~\eqref{eqn:end_conditions_eed} which reads:
\begin{equation}
    \delta\vec{r}^{N-1}=-\sum_{i=1}^{N-2}\delta\vec{r}^i.
\end{equation}
In other words, the change in the joining vector of the last step is
completely determined by the changes in the joining vectors of the other
steps.

The imposed end-to-end rotation condition corresponds to
(see~\eqref{eqn:eer_constraint}):
\begin{equation}
    \delta\mat{\mathcal{D}}^{(1,N)}
    =
    \delta\left(\prod_{i=1}^{N-1}\mat{D}^i(\vec{\psi}^i)\right)
    =
    \sum_{i=1}^{N-1}
    \fp{\mat{\mathcal{D}}^{(1,N)}}{\psi^i_\alpha}\delta\psi^i_\alpha
    =
    \mat{0}.
\end{equation}
This condition can be written as:
\begin{equation}
    \fp{\mat{\mathcal{D}}^{(1,N)}}{\psi^{N-1}_\alpha}\delta\psi^{N-1}_\alpha
    =
    -\sum_{i=1}^{N-2}
    \fp{\mat{\mathcal{D}}^{(1,N)}}{\psi^{i}_\alpha}\delta\psi^{i}_\alpha,
\end{equation}
and with the help of~\eqref{eqn:rotation_product_derivatives} we obtain:
\begin{equation}
    \mat{P}^{N-1}_\alpha\delta\psi^{N-1}_\alpha
    =
    -{\mat{d}^{N-1}}^\T
    \left(\sum_{i=1}^{N-2}\mat{d}^i\mat{P}^i_\alpha{\mat{d}^i}^\T\right)
    \mat{d}^{N-1}\delta\psi^{N-1}_\alpha.
\end{equation}
The right-hand side corresponds to a sum of skew matrices and we therefore
introduce $\mat{Q}^{i,N-1}$ as:
\begin{equation}
    \matelem{
        {\mat{d}^{N-1}}^\T\mat{d}^i\mat{P}^i_\alpha{\mat{d}^i}^\T\mat{d}^{N-1}
    }_{\beta\gamma}\delta\psi^i_\alpha
    =
    -\epsilon_{\beta\gamma\zeta}
    \matelem{\mat{Q}^{i,{N-1}}}_{\zeta\alpha}\delta\psi^i_\alpha.
\end{equation}
Using~\eqref{eqn:step_Xi_matrix} we obtain after contracting the Levi-Civita
symbols:
\begin{equation}
    \matelem{\mat{\Xi}^{N-1}}_{\zeta\alpha}\delta\psi^{N-1}_\alpha
    =
    -\sum_{i=1}^{N-2}
    \matelem{\mat{Q}^{i,{N-1}}}_{\zeta\alpha}\delta\psi^i_\alpha.
\end{equation}
We can invert the left-hand side (see~\eqref{eqn:step_Omega_matrix}) to
finally obtain:
\begin{equation}
    \delta\vec{\psi}^{N-1}
    =
    -\sum_{i=1}^{N-2}
    \mat{\Omega}^{N-1}\mat{Q}^{i,{N-1}}\delta\vec{\psi}^i.
    =
    -\sum_{i=1}^{N-2}\mat{K}^{i}\delta\vec{\psi}^i.
    \label{eqn:matrix_K_definition}
\end{equation}
Note that the matrix $\mat{Q}^{i,{N-1}}$ can be expressed in terms of the
vectors $\vec{S}^i_\beta$:
\begin{equation}
    \matelem{\mat{Q}^{i,{N-1}}}_{\alpha\beta}
    =
    \vec{d}^{N-1}_\alpha\cdot\vec{S}^i_\beta.
\end{equation}
In other words, the matrix $\mat{Q}^{i,{N-1}}$ is obtained by projecting the
vectors $\vec{S}^i_\beta$ in the base-pair frame $\mat{d}^{N-1}$ and, hence,
expresses the infinitesimal rotation originating from the changes in the
angular step dofs $\vec{\psi}^i$ in the base-pair frame $\mat{d}^{N-1}$.

\begin{figure}[htb]
        \center
        \includegraphics[width=.5\textwidth]{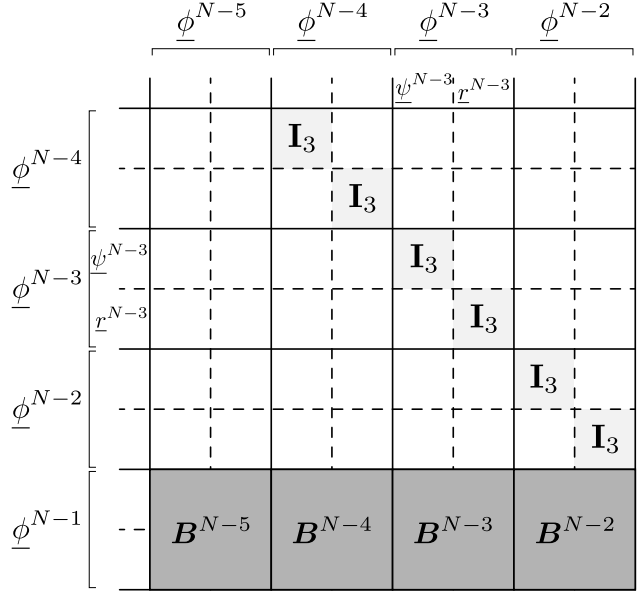}
        \caption{Structure of the matrix $\mat{\hat{J}}$ for imposed
        end-to-end vector and rotation for the last four steps of a collection
        of base pairs. The vectors $\vec{\phi}^i$ denote the step dofs. The
        columns of this matrix are related to the independent step dofs and,
        hence, the last column is for step $(N-2)$ as the $(N-1)$-th step is
        not independent due to the boundary conditions. The solid lines
        delimit the $6\times6$ base-pair step block matrices, while the dashed
        lines separate the angular variables $\vec{\psi}^i$ from the
        translational variables $\vec{r}^i$. The white blocks represent null
        entries.}
        \label{fig:bc_jacobian}
    \end{figure}

The results obtained in~\eqref{eqn:end_conditions_eed}
and~\eqref{eqn:matrix_K_definition} show that:
\begin{equation}
    \delta{\vec{\phi}^{N-1}}
    =
    -\sum_{i=1}^{N-2}
    \left[\:
        \begin{BMAT}(@,20pt,20pt){c:c}{c:c}
            \mat{K}^i & \mat{0} \\
            \mat{0} & \mat{\mathrm{I}}_3
        \end{BMAT}
    \:\right]\delta\vec{\phi}^i
    =
    -\sum_{i=1}^{N-2}\mat{B}^i\delta\vec{\phi}^i.
    \label{eqn:end_conditions_B_matrix}
\end{equation}
This shows that the imposed end-to-end vector and rotation conditions reduce
the number of independent step dofs (the step dofs of the last step can be
expressed in terms of the step dofs of all the other steps).
That is, the set of independent step dofs, denoted $\vec{\hat{\Phi}}$,
corresponds to $\vec{\hat{\Phi}}=\{\vec{\phi}^i\}_{i=1,\dots,N-2}$.
The Jacobian matrix $\mat{\hat{J}}$, defined as
$\mat{\hat{J}}=\partial\vec{\Phi}/\partial\vec{\hat{\Phi}}$, is sparse and its
structure is shown in Fig.~\ref{fig:bc_jacobian}.

\section{Frozen steps}
\label{app:frozen_steps}
We focus in this appendix on the calculation of the matrix $\mat{\tilde{J}}$
(see~\eqref{eqn:frozen_jacobian}).

We consider that the $k$-th step in the collection of base pairs is frozen,
that is, its step parameters $\vec{p}^k$ are imposed and constant. It follows
directly that:
\begin{equation}
    \delta\vec{\psi}^k=\vec{0}.
\end{equation}
The condition on the variation of the translational step dofs,
$\delta\vec{r}^k$, is obtained from the expression:
\begin{equation}
    \delta\vec{\rho}^k
    =
    \fp{\rho^k_\alpha}{r^k_\beta}\delta r^k_\beta
    +
    \sum_{\ell=1}^{k-1}\fp{\rho^k_\alpha}{\psi^\ell_\beta}
    \delta\psi^\ell_\beta
    =
    \vec{0}.
\end{equation}
It follows from~\eqref{eqn:jacobian_matrix_T}
and~\eqref{eqn:jacobian_matrix_U}:
\begin{equation}
    \delta r^k_\beta
    =
    \sum_{\ell=1}^{k-1}
    \matelem{\vec{S}^\ell_\gamma\times\vec{r}^k}_\beta\delta\psi^\ell_\gamma
    =
    \sum_{\ell=1}^{k-1}
    \matelem{\mat{W}^{\ell,k}}_{\beta\gamma}\delta\psi^\ell_\gamma.
    \label{eqn:frozen_vector_r}
\end{equation}
This result simply mean that the vector $\vec{r}^k$ only changes in
orientation due to the changes in the angular step dofs of the preceding
steps.

\begin{figure}[htb]
        \center
        \includegraphics[width=.5\textwidth]{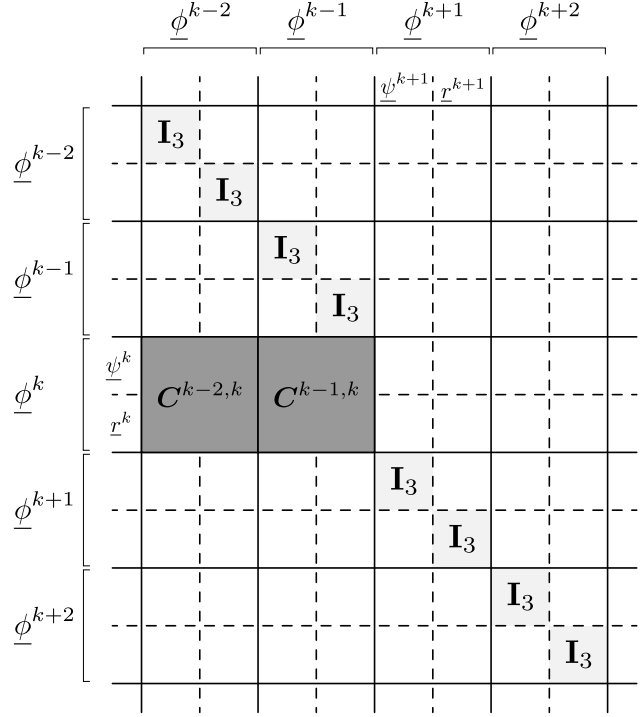}
        \caption{Structure of the matrix $\mat{\tilde{J}}$ for a base-pair
        collection in which the $k$-th step is frozen. The solid lines delimit
        base-pair step block matrices, while the dashed lines separate the
        angular variables $\vec{\psi}^i$ from the translational variables
        $\vec{r}^i$. The white blocks represent null entries.}
        \label{fig:frozen_jacobian}
    \end{figure}

We introduce the matrix $\mat{C}^{\ell,k}$ as:
\begin{equation}
    \delta\vec{\phi}^k
    =
    \sum_{\ell=1}^{k-1}
    \mat{C}^{\ell,k}\delta\vec{\phi}^\ell
    =
    \sum_{\ell=1}^{k-1}
    \left[\:
        \begin{BMAT}(@,20pt,20pt){c:c}{c:c}
            \mat{0} & \mat{0} \\
            \mat{W}^{\ell,k} & \mat{0}
        \end{BMAT}
    \:\right]\delta\vec{\phi}^\ell.
    \label{eqn:frozen_jacobian_matrix_C}
\end{equation}
The Jacobian matrix $\mat{\tilde{J}}$ is readily determined from this result
and its structure is shown in Fig.~\ref{fig:frozen_jacobian}.

\section{Zajac instability for DNA minicircles}
\label{app:zajac_instability}
The Zajac instability~\cite{Zajac:Stability-of-two-planar-l:1962,
Goriely:Twisted-Elastic-Rings-and:2006} is related to the fact that a twisted
elastic ring becomes unstable if the twist density within the ring exceeds a
certain value.
It is straightforward to transpose this result to our DNA base-pair model (in
fact, our mesoscale model together with our ideal force field can be seen as a
discretized Kirchhoff elastic rod model).

For a DNA fragment of $N$ bp in its rest state, the intrinsic total twist
$\overline{\mathrm{Tw}}$ is given by:
\begin{equation}
    \overline{\mathrm{Tw}}=\frac{N*\overline{\theta}_{3}}{360}=\frac{N}{10.5},
\end{equation}
where $\overline{\theta}_{3}=360/10.5=34.2857~\deg$ is the intrinsic twist
step parameter value.
We introduce the excess of twist
$\Delta\mathrm{Tw}=\mathrm{Tw}-\overline{\mathrm{Tw}}$ and the stability
criterion is written as:
\begin{equation}
    \left|\Delta\mathrm{Tw}\right| < \frac{\sqrt{3}}{2\Gamma},
\end{equation}
where $\Gamma$ is the ratio of the bending and twisting stiffnesses (for our
ideal for field we have $\Gamma=0.715$). Since we only consider the case of a
planar minicircle we have
$\mathrm{Tw}=\mathrm{Lk}=\Delta\mathrm{Lk}+\mathrm{Lk}^0$ and we recall that
$\mathrm{Lk}^0=\left[N/10.5\right]$ (where the brackets stand for the nearest
integer operator). It follows that the criterion can be written as:
\begin{equation}
    -\frac{\sqrt{3}}{2\Gamma}
    - \left(\left[\frac{N}{10.5}\right] - \frac{N}{10.5}\right)
    <
    \Delta\mathrm{Lk}
    <
    \frac{\sqrt{3}}{2\Gamma}
    - \left(\left[\frac{N}{10.5}\right] - \frac{N}{10.5}\right).
\end{equation}
These conditions with the fact that $\Delta\mathrm{Lk}$ is an integer leads
to the following set of solutions:
\begin{equation}
    \Delta\mathrm{Lk} = \left\{-1,0,+1\right\}.
\end{equation}

\section{Implementation details}
\label{app:implementation}
We use the C{}\verb!++! ALGLIB library~\cite{Bochkanov:ALGLIB:} to implement
our minimization method and rely on the L-BFGS algorithm.
We also use the C{}\verb!++! Eigen library~\cite{Guennebau:Eigen-v3:2010} to
handle the various matrix computations and manipulations.
Our minimization software is available upon request.

\end{document}